\documentclass[prd,letterpaper,twocolumn,tightenlines,superscriptaddress,showpacs,preprintnumbers,nofootinbib,floatfix,psfig,showpacs,showkeys]{revtex4-1}

\usepackage{amsmath,amssymb,amsfonts,graphicx,pifont,dcolumn}
\usepackage{epstopdf}
\usepackage{color}
\usepackage{cancel}
\usepackage{ulem}
\RequirePackage{slashed}

\def\xP{x_{\!I\!\!P}}

\def\bmDelta{\mbox{\boldmath$\Delta$}}

\newcommand{\bit}{\begin{itemize}}
\newcommand{\eit}{\end{itemize}}

\newcommand{\etal}{{\it et al.}}

\def\be{\begin{equation}}
\def\ee{\end{equation}}
\def\bea{\begin{eqnarray}}
\def\eea{\end{eqnarray}}

\begin{document}




\twocolumngrid

\title{ Diffractive incoherent vector meson production off protons: \\ a quark model approach to gluon fluctuation effects}

\author{Marco~Claudio~Traini} 
\affiliation{INFN - TIFPA, Via Sommarive 14, I-38123 Trento-Povo, Italy}
\affiliation{Dipartimento di Fisica, Universit\`a degli Studi di Trento, 
Via Sommarive 14, I-38123 Trento-Povo, Italy}
\author{Jean-Paul~Blaizot}
\affiliation{Institut de Physique Th\'eorique, Universit\'e Paris Saclay, CEA, F-91191 Gif-sur-Yvette, France}


\begin{abstract}
Fluctuations play an important role in diffractive production of vector mesons. It was in particular recently suggested, based on the Impact-Parameter dependent Saturation model (IPSat),  that geometrical fluctuations triggered by the motion of the constituent quarks within the protons could explain incoherent diffractive processes observed at 
HERA. We propose a variant of the IPSat model which includes spatial and symmetry correlations between constituent quarks, thereby reducing the number of parameters needed to describe diffractive vector meson production to a single one, the size of the gluon cloud around each valence quark. The application to $J/\Psi$, $\rho$ and $\phi$ diffractive electron and  photon production cross sections reveal the important role of geometrical fluctuations in incoherent channels, while other sources of fluctuations are needed to fully account for electroproduction of light mesons, as well as photo production of $J/\Psi$ mesons at small momentum transfer. 
\end{abstract}

\maketitle

\section{\label{intro}introduction}

Fluctuations play an essential role in the diffractive production of vector mesons. It was recently suggested that these fluctuations could be dominated by those, event by event, of the constituent quark positions inside the proton, and that these could be constrained by the incoherent diffractive photoproduction of $J/\Psi$ mesons off protons \cite{M&Schenke2016}. Such fluctuations, of essentially geometrical origin, are commonly referred to as ``geometrical fluctuations''. They are the analog of the fluctuations linked to the positions of the nucleons in high energy nucleus-nucleus collisions \cite{Miller:2007ri}.  

As we shall see, a crucial ingredient entering the calculation of the diffractive processes is the cross section of a small color dipole crossing the proton at a given impact parameter. The interaction of the dipole with the proton is directly sensitive to the total density of gluons that it ``sees'' on its path through the proton. Although we have experimental information about the total (integrated over the impact parameter) density of gluons in a proton, the dependence on the impact parameter is much less constrained. 
A simple dipole model that includes the physics of saturation and takes explicitly into account  the impact parameter dependence of gluon distributions is the Impact-Parameter dependent Saturation model (IPSat) \cite{IPSat2002,IPSat2003,IPSat2006,IPSat2013,IPSat2017}. 
In the IPSat model the impact parameter dependence of the amplitude is simple to implement and it can be easily generalized from Deep Inelastic Scattering (DIS) off protons to DIS off nuclei \cite{DIS1,DIS2,DIS2017,DIS3}. Other excellent probes of the high energy saturation regime are the exclusive diffractive processes in the electron-proton collisions: exclusive vector meson production and deeply virtual Compton Scattering (DVCS) are the prominent examples. 

Our main interest, in the present work, is the physics of exclusive diffractive meson production, since an interesting new piece of information can be extracted from such reactions, namely how much the spatial gluon distribution fluctuates, event-by-event, within a proton. Experimentally one can access this information via exclusive {\it incoherent} diffractive meson production, i.e. events connected with a dissociated proton \cite{M&Schenke2016PRL}. Including the analysis of {\it coherent} diffractive processes where the proton remains intact, both the impact parameter dependence and the fluctuations of the gluon distribution in the proton can be constrained \cite{M&Schenke2016}.  Different final states depend in different ways on the impact parameter,  where intrinsically non-perturbative physics may become relevant. Thus the fluctuations of the shape of the gluon distribution may be influenced by non-perturbative physics and the aim of the present work is a detailed study of some of such non-perturbative effects. To this end we  present 
a self-consistent approach where the spatial quark  and gluons distributions are consistently  calculated. The number of parameters drastically reduces and the predictive power of the IPSat model increases since it is based on calculated properties of the quark wave functions. \\

The paper is organized as follows. 
In Sect.~\ref{sec:V-DIS} we review the approach used by M\"antysaari and Schenke in Ref.~\cite{M&Schenke2016} to calculate the vector meson production cross sections. In particular,  we stress the role of geometrical fluctuations in the description of incoherent photoproduction of $J/\Psi$ mesons.
In Sect.~\ref{sec:qcorrelations} we present and discuss the quark correlations which are relevant in the description of diffractive processes. These are obtained in a specific quark model that allows for a simple determination of the quark wave functions of the nucleons. Fluctuations in the density of gluons are introduced, as in Ref.~\cite{M&Schenke2016}, by attaching a gluon cloud around each valence quark. 
In Sect.~\ref{sec:Q_P} we use DGLAP evolution equations to various degrees of precision in order to relate quarks, gluons and sea quark degrees of freedom at the initial non-perturbative scale to their values at the large experimental scale.  
In Sec. \ref{sec:2hoJ_Psi} we present our main results for the coherent and incoherent $J/\Psi$ photoproduction, while $J/\Psi$, $\rho$ and $\phi$ electron photoproduction is discussed in Sect.~\ref{sec:electroproduction}. 
Finally, conclusions are drawn in Sect.~\ref{sec:conclusions}.

\section{\label{sec:V-DIS}Diffractive Deep Inelastic Scattering in the dipole picture}

In deep inelastic lepton-proton scattering, the exclusive production of vector mesons ($V$) proceeds via the exchange of pomerons  in the case of a {\it diffractive} process where no color is exchanged between the proton and the produced system. The absence of colored strings leads to a rapidity gap (a region in rapidity with no produced particles) which characterizes experimentally the diffractive events.
If the scattered proton remains intact, the process is called {\it coherent}, while for {\it incoherent} processes the final proton breaks up (see Ref.~\cite{BP2002} for an introduction to diffractive processes and their description within perturbative QCD).

 Explicitly, using the notation of Ref.~\cite{M&Schenke2016}, we write the {\it coherent} diffractive cross section as
\be
{d \sigma_{T,L}^{\gamma^* p \to V p} \over dt} = {(1+\beta^2) \over 16 \pi}
\left|\langle {\cal A}_{T,L}^{\gamma^* p \to V p}(x_{\xP},Q^2, \bmDelta)\rangle\right|^2\,,
\label{eq:dxsection_coh}
\ee
where ${\cal A}_{T,L}^{\gamma^* p \to V p}(x_{\xP},Q^2, \bmDelta)$ is the scattering amplitude, ${\xP}=(P-P')\cdot q/(P\cdot q)$ the fraction of the longitudinal momentum of the proton transferred to the pomeron ($\!I\!\!P$),  and the momentum transfer (square) is $t = -(P'-P)^2$ with $P$ and $P'$ the initial and final proton four-momenta. The virtual photon-proton scattering is characterized by a total center-of-mass-energy squared $W^2 = (P+q)^2$, ($Q^2 = -q^2$). Finally ${\bmDelta}=(P'-P)_\perp$ is the transverse momentum transfer\footnote{Throughout this paper we use bold face letters to denote vectors in the transverse plane.}.

The amplitude ${\cal A}_{T,L}^{\gamma^* p \to V p}(x_{\xP},Q^2, \bmDelta)$ for diffractive vector meson production assumes the form \cite{IPSat2003,IPSat2006}
\bea
&&{\cal A}_{T,L}^{\gamma^* p \to V p}(x_{\xP},Q^2, \bmDelta) = i \int d^2 {\bf r} \int d^2 {\bf b} \int {dz \over 4 \pi}  \nonumber \\
&\times& \left(\Psi^* \Psi_V \right)_{T,L}(Q^2,{\bf r},z) \, e^{-i[{\bf b}-(1-z) {\bf r}] \cdot {\bmDelta}}  \nonumber \\
&\times&{{d \sigma_{q \bar q} \over d^2 {\bf b}}}({\bf b},{\bf r}, x_{\xP}),
\label{eq:A_exclusive}
\eea
where the subscripts $T$ and $L$ refer to transverse and longitudinal polarization of the exchanged virtual photon.
This expression  is based on the dipole picture:  the photon fluctuates into a  quark-antiquark pair, a color dipole, with  transverse size  ${\bf r}$, while $z$ is the fraction of the photon's light-cone momentum carried by a quark. This picture holds in a frame where  the dipole lifetime is much longer than the interaction time with the target proton. The $\gamma^* p$  scattering then proceeds through three steps: i) The incoming virtual photon fluctuates into a quark - antiquark pair; the splitting of the photon is described by the virtual photon wave function $\Psi$, which can be calculated in perturbative QED (see e.g. Ref.~\cite{gammaQED1}). ii) The $q$-$\bar q$ pair scatters on the proton, with  a  cross section $\sigma_{q \bar q}$ to be discussed below. This cross section is Fourier transformed into momentum space,  with  the transverse momentum transfer  $\bmDelta$ conjugate to ${\bf b}-(1-z) {\bf r}$ (distance, in the transverse plane, from the center of the proton to the center-of-mass of the dipole \cite{IPSat2006}). iii) The scattered dipole recombines to form a final state, in the present case the vector meson with wave function $\Psi_V$ (cf.~\ref{sec:overlapPSI}). The factor $(1+\beta^2)$ in Eq.~(\ref{eq:dxsection_coh}), is described in~\ref{sec:phen_corrections} together with other phenomenological corrections. 

In Eq.~(\ref{eq:dxsection_coh}) the amplitude is averaged over the proton ground state, as indicated by the angular brakets. When breakup processes are included, the square of the average amplitude leaves the place to a sum over intermediate states. Ignoring in that sum the contribution of the ground state, which yields the coherent part of the cross section, we are left with the {\it incoherent}  cross section. This takes the form \cite{M&Schenke2016}
\bea
&& {d \sigma_{T,L}^{\gamma^* p \to V p'} \over dt} = {(1+\beta^2) \over 16 \pi}\left[ \langle\left| {\cal A}_{T,L}^{\gamma^* p \to V p}(x_{\xP},Q^2, \bmDelta)\right|^2 \rangle \right.\nonumber \\
&& - \left. \left| \langle {\cal A}_{T,L}^{\gamma^* p \to V p}(x_{\xP},Q^2, \bmDelta)\rangle\right|^2\right]
\label{eq:dxsection_incoh}
\eea
and involves the variance of the amplitude.  Note that, as written in Eq.(\ref{eq:A_exclusive}), the amplitude ${\cal A}$ is averaged over the dipole size and the impact parameter. 

\subsection{\label{sec:CoherentGaussian}Coherent production}
\begin{figure}[tbp]
\centering\includegraphics[width=\columnwidth,clip=true,angle=0]{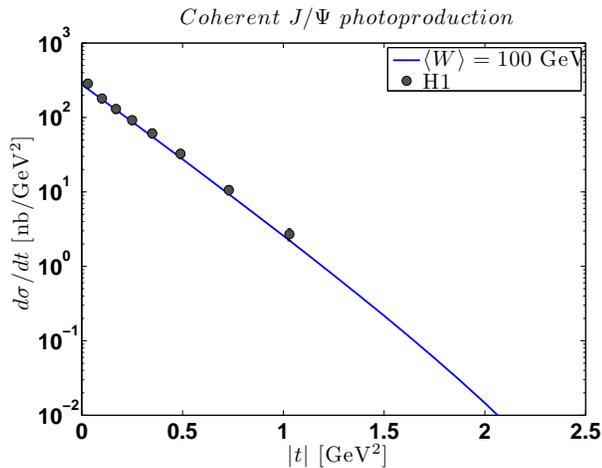}
\caption{\small (color on line) Coherent photoproduction ($Q^2 = 0$) cross section within the kinematical conditions of the HERA experiments:
$\xP \approx 9.6 \cdot 10^{-4}$ for $\langle W\rangle = 100$ GeV.
The parameter of Eq.~(\ref{eq:TpGaussian}) is chosen to be $B _G= 4$ GeV$^{-2}$ . The H1 data are from \cite{H1W100coh2006,H1W75coh2013}.}
\label{fig:HERA_J_PSI_photo1}
\end{figure}
In this paper, we shall rely on the
the IPSat model \cite{IPSat2002,IPSat2003,IPSat2006,IPSat2013,IPSat2017}, which has been very successful in describing a wide range of data from HERA.
In this model the dipole cross section is given by (see e.g. \cite{Mueller1990})
\be
{d \sigma_{q \bar q} \over d^2 {\bf b}} = 2 \left[1-\exp\left(- {\pi^2 \over 2 N_c} {\bf r}^2 \alpha_S(\mu^2)\,{\xP} g({\xP},\mu^2)\,T({\bf b}) \right)\right]
\label{eq:dXsqbarq}
\ee
where the proton (transverse) spatial profile function $T_G({\bf b})$ is assumed to be Gaussian in a first approximation, viz.
\be
T({\bf b}) = T_G({\bf b}) = {1 \over 2 \pi B_G} e^{-{{\bf b}^2 / (2 B_G)}}. \label{eq:TpGaussian}
\ee
The scale $\mu$ in the gluon distribution function ${\xP} g({\xP},\mu^2)$ is related to the size ${\bf r}$ of the dipole
\be
 \mu^2 = \mu^2({\bf r}^2) = \mu_0^2 + {4 \over {\bf r}^2},\label{eq:mu} 
\ee
and the gluon distribution is parameterized as 
\be
 x g(x,\mu_0^2) = A_g \, x^{-\lambda_g}\,(1-x)^{5.6}. \label{eq:xg}
\ee
Loosely speaking, what the IPSat model does in Eq.~(\ref{eq:dXsqbarq}),  is to take the integrated gluon distribution (\ref{eq:xg}), and redistribute the gluons in transverse plane according to the phenomenological profile $T({\bf b})$ given in Eq.~(\ref{eq:TpGaussian}).
\begin{figure}[tbp]
\centering\includegraphics[width=\columnwidth,clip=true,angle=0]{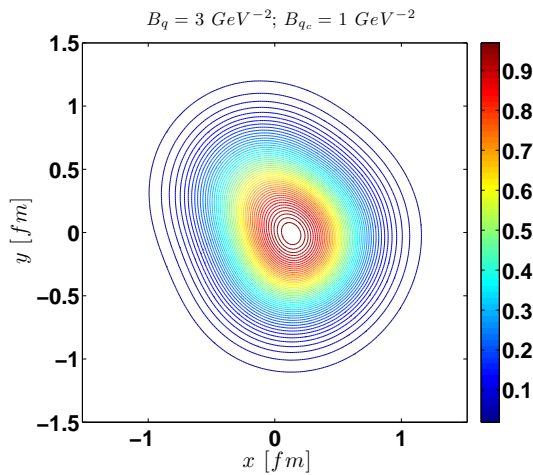}
\caption{\small (color on line) Example of a smooth density profile obtained from Eq.~(\ref{eq:Tbmodified}) with the parameters indicated in the panel.}
\label{fig:shapes2}
\end{figure}

As an illustration of the results obtained within such an approach, we display in  Fig.~\ref{fig:HERA_J_PSI_photo1} the cross section for the coherent $J/\Psi$ diffractive photoproduction ($Q^2 = 0$, real photons,  and therefore transverse response only). The results shown in Fig.~\ref{fig:HERA_J_PSI_photo1} reproduce those shown in Fig. 6  of Ref.~\cite{M&Schenke2016}, for $B=4$ GeV$^{-2}$. The scale $\mu_0^2$ entering the initial condition  for the DGLAP evolution of the gluon distribution ${\xP} g({\xP},\mu^2)$ \cite{IPSat2002}, is taken from Ref.~\cite{IPSat2013} ($m_c = 1.4$ GeV is used for the charm quark mass). 

\subsection{\label{sec:Gaussianfluctuations} Incoherent $J/\Psi$ diffractive production}

\begin{figure}[tbp]
\centering\includegraphics[width=\columnwidth,clip=true,angle=0]{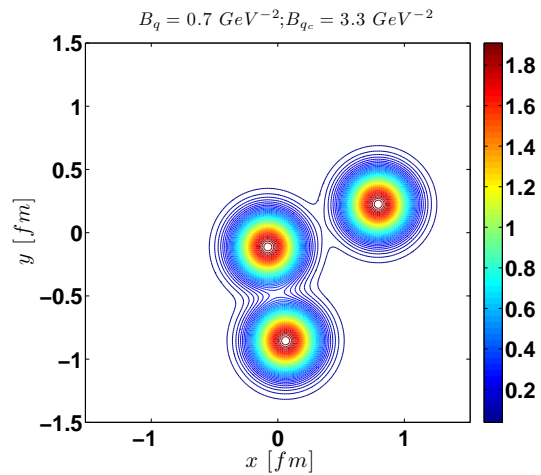}
\caption{\small (color on line) Examples of a ``lumpy" density profile obtained from Eq.~(\ref{eq:Tbmodified}) with the parameters indicated in the panel.}
\label{fig:shapes1}
\end{figure}

The {\it incoherent} component of the diffractive cross section for vector meson production involves the fluctuation of the amplitude (see Eq.~(\ref{eq:dxsection_incoh})). Following the authors of Ref.~\cite{M&Schenke2016}, we  assume that these fluctuations have  a geometrical origin, i.e., they  are dominated by the fluctuations, event by event, of the locations of the constituent quarks in the transverse plane. 
We then  consider  the   density $T({\bf b})$ in Eq.~(\ref{eq:dXsqbarq})  as resulting from the sum of the contributions of the individual quarks, i.e., 
\be
T({\bf b}) \to {1 \over N_q} \sum_{i=1}^{N_q}T_q({\bf b}-{\bf b}_i),
\label{eq:Tbmodified}
\ee
with 
\be
T_q({\bf b}) = {1 \over 2 \pi B_q}e^{-{\bf b}^2/(2 B_q)}
\label{eq:Tq}
\ee
with parameter $B_q$. That is, we assume that   each constituent quark is surrounded by a cloud of gluons, assumed also to be Gaussian, and represented by $T_q$ in Eq.~(\ref{eq:Tq}). 

In practice one starts sampling the constituent quarks' positions  in the transverse plane (${\bf b}_i$, $i=1,2,3$), from a Gaussian distribution with width parameter $B_{qc}$, neglecting any possible correlations between the quarks \cite{M&Schenke2016}. 
For fixed $N_q$ ($N_q = 3$) the degree of fluctuations is controlled by the relative sizes of the parameters $B_{qc}$ and $B_q$. In Fig.~\ref{fig:shapes1} an example of a ``lumpy" proton configuration is shown: it corresponds to a relatively broad distribution of constituent quarks, $B_{qc}= 3.3$ GeV$^{-2} = (0.3585\,{\rm fm})^2$, and a small size gluon cloud around each valence quarkx,  $B_q = 0.7$ GeV$^{-2} = (0.1651\,{\rm fm})^2$. In contrast, 
Fig.~\ref{fig:shapes2} shows a ``smooth" proton that has little fluctuations: this corresponds to a compact distribution of constituent quarks, $B_{qc}= 1.0$ GeV$^{-2} = (0.1973\,{\rm fm})^2$, with a broad distribution of gluons around each constituent quark, $B_q = 3.0$ GeV$^{-2} = (0.3418\,{\rm fm})^2$). The parameters are chosen in such a way that the two-dimensional gluon root mean square radius of the proton is kept at the fixed value 
\bea
\sqrt{\langle {\bf b}^2\rangle} & = & \sqrt{2 B} = \sqrt{2(B_{qc}+B_q)} = \nonumber \\
& = & 2 \sqrt{2}\; {\rm GeV^{-1}} \approx 0.55\; {\rm fm}. 
\label{eq:BqcBq}
\eea

\begin{figure}[tbp]
\centering\includegraphics[width=\columnwidth,clip=true,angle=0]{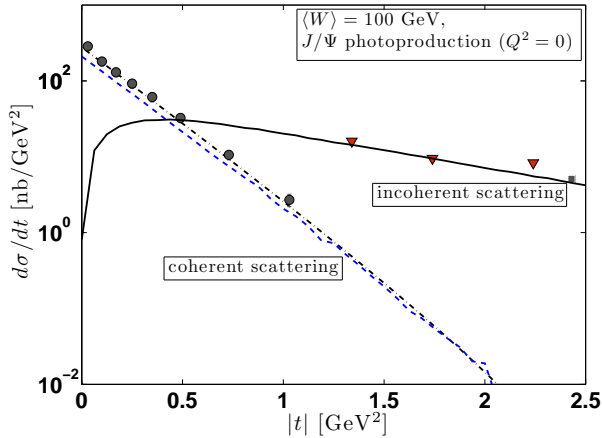}
\caption{\small (color on line) Coherent and incoherent $J/\Psi$ photoproduction cross section at $\langle W\rangle =100$ GeV. 
The full line show the results for the incoherent scattering and the lumpy configuration parameters of  Fig.~\ref{fig:shapes1} ($B_{qc} = 3.3$ GeV$^{-2}$, $B_q=0.7$ GeV$^{-2}$). The same parameters used for the coherent scattering give the result shown by the dashed line. For comparison also the coherent results obtained  without geometric fluctuations ($B=B_G=4$ GeV$^{-2}$ (dot-dashed line)) are shown.
Coherent H1 data from refs.\cite{H1W100coh2006} and \cite{H1W75coh2013} (circles),  incoherent data (triangles) from H1 and ZEUS experiments of refs.\cite{H1W75coh2013} and \cite{ZEUSincoh2003}. The single square refers to the total cross section of the H1 experiment at large momentum transfer~\cite{H1total2003}.}
\label{fig:HERA_J_PSI_photo2}
\vspace{-1.0em}
\end{figure}
The configurations obtained via the sampling procedure just described represent the basic ingredients for a  complete calculation of the {\it coherent} and {\it incoherent} diffractive vector meson production. The number of configurations considered in the present study for the evaluation of Eqs.~(\ref{eq:dxsection_coh}) and (\ref{eq:dxsection_incoh}), is $N_{\rm conf}=10000$. We have checked that the results of our simulations are stable when $N_{\rm conf}$ is increased beyond this value. 
We show in Fig.~\ref{fig:HERA_J_PSI_photo2} the results obtained for the  photoproduction cross sections, in the kinematical conditions of the HERA experiments, and for the ``lumpy" configurations of Fig.~\ref{fig:shapes1}.
As can be seen the coherent as well as incoherent data of  the HERA experiments are well reproduced (see the captions of Fig.~\ref{fig:HERA_J_PSI_photo2}  for more details). For comparison also the coherent results obtained  without geometric fluctuations and an average Gaussian profile (with $B = B_G = 4$ GeV$^{-2}$) are shown (cf. Eq.~(\ref{eq:TpGaussian}) and Fig.~\ref{fig:HERA_J_PSI_photo1}).

\begin{figure}[tbp]
\centering\includegraphics[width=\columnwidth,clip=true,angle=0]
{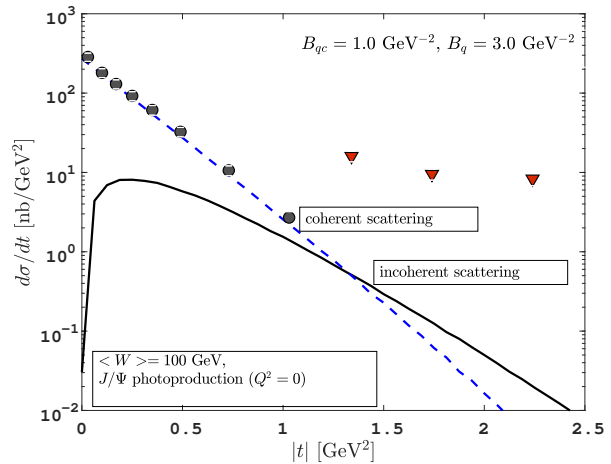}
\caption{\small (color on line) Coherent and incoherent $J/\Psi$ photoproduction cross section at $\langle W\rangle =100$ GeV.  The full line shows the result for the incoherent scattering using the ``smooth" configuration parameters   (see legend and Fig.~\ref{fig:shapes2}). Data as in Fig.~\ref{fig:HERA_J_PSI_photo2}. }
\label{fig:HERA_J_PSI_photosmooth}
\end{figure}

Finally, we consider the respective influence of  the fluctuations on coherent and incoherent cross sections. If one smoothens the strength of the fluctuations by choosing as Gaussian parameters the values of   Fig.~\ref{fig:shapes2} (i.e.  $B_{qc} = 1.0$ GeV$^{-2}$ and $B_{q} = 3.0$ GeV$^{-2}$) and calculates again coherent and incoherent cross sections for diffractive photon-production at HERA kinematical conditions,  the results of Fig.~\ref{fig:HERA_J_PSI_photosmooth} are obtained.  The incoherent cross section is largely underestimated, while the calculated coherent cross section  reproduces the HERA data. This just confirms the conclusion of Ref.~\cite{M&Schenke2016} regarding the sensitivity of the incoherent scattering to the strength of the  (geometrical) gluon fluctuations.

\section{\label{sec:QMBApproach}A quark model based approach to diffractive scattering}

The description of incoherent diffractive vector meson production
that has been discussed in the previous section relies on simple Gaussian approximations for the quark distribution as well as the gluon distribution around each constituent quark. They have revealed the large  sensitivity of the process to the  fluctuations in these distributions.  However the calculation, which essentially duplicates that of Ref.~\cite{M&Schenke2016},  completely neglects correlations between the constituent quarks. Such correlations could however affect the gluon fluctuations. Our goal in the next sections is to develop a simple treatment of these correlations, based on a quark model for the nucleon wave function (QMBA). As an outcome of this approach, we shall see that the number of free parameters to describe the diffractive scattering is drastically reduced and the predictions are more directly related to the quark and parton dynamics.

The correlations among (constituent) quarks are induced by their mutual interaction, in particular by the One-Gluon-Exchange. In the non-relativistic limit, this yields the so-called hyperfine interaction \cite{OGEx}
\bea
&&V_{hyp} = {2 \over 3}{\alpha_S \over m^2} \sum_{i < j} {8 \pi \over 3} \left\{ \vec S_i \cdot \vec S_j \, \delta(\vec r_{ij}) + \right.\nonumber \\
&& + \left.{1 \over r_{ij}^2} \left[ 3 \,(\vec S_i \cdot \hat r_{ij}) \, (\vec S_j \cdot \hat r_{ij}) - \vec S_i \cdot \vec S_j\right]\right\}\label{Hyp}\,.
\eea
This interaction introduces a spin dependence in the quark wave function. In particular the contact term of Eq.~(\ref{Hyp}) (which is the most relevant) is repulsive in $S = 1$ states ($u u$ pairs in protons and $d d$ pairs in neutrons) and attractive in $S = 0$ ($u d$ pairs). It contains also a tensor component expressed in terms of the quark spin $\vec S_i$ and the relative coordinates $\vec r_{ij}$.
The $N - \Delta$ mass difference (fixed at about $300$ MeV) also fixes the value of $\alpha_S$ (see e.g. Ref.\cite{GianniniSantopinto2015}).

\subsection{The Isgur and Karl model and SU(6) breaking} \label{sec:qcorrelations}

The presence of the hyperfine interaction naturally breaks $SU(6)$ symmetry and leads to a description of the proton as a superposition of different $SU(6)$ configurations (multiplets 56, 70). A specific realization  is given by the model introduced by Isgur and Karl, where, by diagonalizing the Hamiltonian in a harmonic oscillator (h.o.) basis up to $2 \hbar \omega_0$ states, one finds the following  nucleon wave function \cite{IsgurKarl1979} 
\bea
|N\rangle & = & a_S |56, N\, ^2S_{1/2}\rangle_S + a'_{S} |56',N\, ^2S'_{1/2}\rangle_M + \nonumber \\
& + & a_M |70, N\, ^2S_{1/2}\rangle_M + a_D |70, N\, ^4D_{1/2}\rangle_M\,.\label{IK_parameters}
\eea
The first state in Eq.~(\ref{IK_parameters}) is in the
$0\hbar \omega_0$-shell, while the remaining ones are all $2\hbar \omega_0$ states. 
The explicit values of the parameters obtained by Isgur and Karl are
$$
a_S = 0.931,\; a'_S = - 0.274,\; a_M = - 0.233,\; a_D = -0.067\,.
$$
Neglecting the breaking of the $SU(6)$ symmetry would give  $a'_S = a_M = a_D = 0$ and the spatial distributions of the $u$ and $d$ valence quarks cannot reproduce the charge distribution in the neutron.  

\subsection{\label{sec:T2ho}The two-harmonic-oscillator (2 h.o.) model}

In the present study we simplify the picture with in mind  the description of the scattering properties. We will describe the $SU(6)$-breaking effects induced  by the hyperfine interaction within a harmonic-oscillator model and introduce two different force constants between $u$ and $d$ quarks. For the nucleons, the procedure is as follows
(see ref.~\cite{ConciTraini1990}): the nucleons $p$ and $n$ are constructed from the two types of constituent quarks, $u$ and $d$, which are considered to be distinct and not to be permuted. The internal quark wave functions are written as $p (uud)$ and $n (ddu)$, in each case taking the first two quarks to be identical. Given spin-dependent forces, the third (unlike) quark will have a different interaction with the first two (like) quarks than these two will have with each other. The justification for using these wave functions has been discussed in detail by Franklin~\cite{Franklin1968} many years ago, and applied by Capstick and Isgur~\cite{CapstickIsgur1986} to construct a relativized quark model for baryons. The two-body potential takes the form
\bea
V & = & {1 \over 2} K \vec r\,^2_{12} + {1 \over 2} K' \left( \vec r\,^2_{13} + \vec r\,^2_{23} \right) =  \nonumber \\
& = & {1 \over 2}\left(2 K + K'\right) \vec \varrho\,^2 + {3 \over 2}K' \vec \lambda\,^2\,,  \label{2Vho}
\eea
where $\vec \varrho = \left(\vec r_1 - \vec r_2 \right)/\sqrt{2}$ and $\vec \lambda = \left(\vec r_1 + \vec r_2 - 2 \vec r_3 \right)/\sqrt{6}$ are Jacobi coordinates. 
Two h.o. constants can be defined:
\bea
\alpha^2 & = & m \omega_{0 \varrho} = \left(m (2K + K')\right)^{1/2}\,,\nonumber \\ 
\beta^2 & = &   m \omega_{0 \lambda} = \left(3mK'\right)^{1/2}\label{alpha-beta}\,.
\eea
The three-quark wave function is then written as
\bea
\Psi_{3q} & = & \left[\Phi_{\rm color}\right]_A \times  {1 \over \sqrt 2} \, \left(\phi_N \times \chi_{M A} + \phi_N \times \chi_{M S}\right) \times \nonumber \\
&\times & {\alpha^{3/2}\,\beta^{3/2} \over \pi^{3/4}}e^{-(\alpha^2 \vec \varrho\,^2 + \beta^2 \vec \lambda\,^2)/2 }\,, \label{eq:2hoNwf}
\eea
where 
$\phi_p=|uud\rangle$, $\phi_n = |ddu\rangle, $
and
$$\chi_{MA} = (\uparrow \downarrow - \downarrow \uparrow) \uparrow /\sqrt 2,\qquad
\chi_{MS} = (\uparrow \downarrow + \downarrow \uparrow) \uparrow /\sqrt 2, $$ 
are the spin components.

A physically sensible way to fix the parameters (\ref{alpha-beta}) is to relate them to the charge r.m.s radius of the proton and neutron:
\bea
\langle r^2 \rangle_p & = & \phantom{+} \left(0.862 \pm 0.012\right)^2\,{\rm fm}^2 = {1 \over \alpha^2}\,, \nonumber 
\\
\langle r^2 \rangle_n & = & - \left(11.94 \pm 0.18\right) \cdot 10^{-2}\,{\rm fm}^2 = - {1 \over 2}\left({1 \over \alpha^2} - {1 \over \beta^2}\right)\,, \nonumber \\
\label{eq:rprn}
\eea
and consequently $\alpha^2 \approx 1.35$ fm$^{-2}$ and $\beta^2 \approx 1.99$ fm$^{-2}$ (corresponding to $K'/K \approx 5.3$). The neutron charge distribution can be reproduced by breaking the $SU(6)$ symmetry ($\alpha \neq \beta$) and vanishes in the $SU(6)$-symmetric limit of a single harmonic oscillator potential ($\alpha = \beta$).

\subsection{\label{sec:T2hob}The density profile function and sampling procedure}

The 2 h.o. average transverse profile function is obtained from the spherical density $\rho_{2ho}(r)$,
\be
T_{2ho}({\bf b}) = {1 \over {\cal N}_u+{\cal N}_d}\int_{-\infty}^{+\infty} dz \, \rho_{2ho}\left(r=\sqrt{z^2+{\bf b}^2}\right),
\ee
where (from the wave function (\ref{eq:2hoNwf}))
\be
\rho_{2ho}(r) = {\cal N}_u  {\kappa_u^3 \over \pi^{3/2}}\,e^{-\kappa_u^2 r^2}+{\cal N}_d  {\kappa_d^3 \over \pi^{3/2}}\,e^{-\kappa_d^2 r^2},
\ee
and $\int d {\bf r}\, \rho_{2ho}(r) = {\cal N}_u+{\cal N}_d = 3$. Consequently 
\bea
T_{2ho}({\bf b}) & = & {{\cal N}_u \over {\cal N}_u+{\cal N}_d}{1\over 2 \pi B_u}\,e^{-{\bf b}^2/(2 B_u)} + \nonumber \\
& + & {{\cal N}_d \over {\cal N}_u+{\cal N}_d}{1 \over 2 \pi B_d}\,e^{-{\bf b}^2/(2B_d)},
\label{eq:T2ho}
\eea
with 
\bea
&& \int d {\bf b}\, T_{2ho}({\bf b})  = 1, \nonumber \\
&& {1 \over 2 B_u} = \kappa_u^2 = {3 \over 2} 4{\alpha^2 \beta^2 \over 3 \alpha^2 + \beta^2} \approx 2.67\;{\rm fm^{-2}};\;\;\;B_u \approx 4.8\;{\rm GeV^{-2}},\nonumber \\
&& {1 \over 2 B_d} = \kappa_d^2 = {3 \over 2}{\beta^2} \approx 2.99\;{\rm fm^{-2}}; \;\;\; B_d \approx 4.3\;{\rm GeV^{-2}}.
\label{eq:BuBd}
\eea

Eq.~(\ref{eq:T2ho}) explicitly summarizes the effects on the profile function of the correlations between quarks that are due to the $SU(6)$-breaking component of the One-Gluon-Exchange and the spin-isospin symmetries of the proton wave function. 
The $SU(6)$-symmetric limit of a single harmonic oscillator wave function is recovered for 
\be
\alpha^2=\beta^2 \to {1 \over 2B_u} = {1 \over 2 B_d} = {3 \over 2}\alpha^2= {1 \over 2 B_0},
\ee 
in which case 
\be
T_{2ho}({\bf b}) \to T_{ho}({\bf b}) =  {1 \over 2 \pi B_0}\,e^{-{\bf b}^2/(2 B_0)},
\label{eq:T2hoTho}
\ee 
i.e.  a Gaussian approximation with $B_0 = 6.34$ GeV$^{-2}$.

The sum in Eq.~(\ref{eq:T2ho}) can be sampled by a random selection of the single term of the sum, followed by sampling the distribution of that term \cite{Kalosetal_2008}. In this way the  correlated positions of the quarks relative to the origin, ${\bf b}_i$ ($i=1,2,3$) are sampled from the 2 h.o. distribution (\ref{eq:T2ho}). 

We should emphasize here that the sampling of the one-body density takes into account the correlations among the constituent quarks only to the extent that these modify the one-body density. In principle, since the full wave-function is known, it should be possible to calculate more fully the effect of these correlations, but this is beyond the scope of the present paper.

\begin{figure}[tbp]
\vspace{-1.0em}
\centering\includegraphics[width=\columnwidth,clip=true,angle=0]{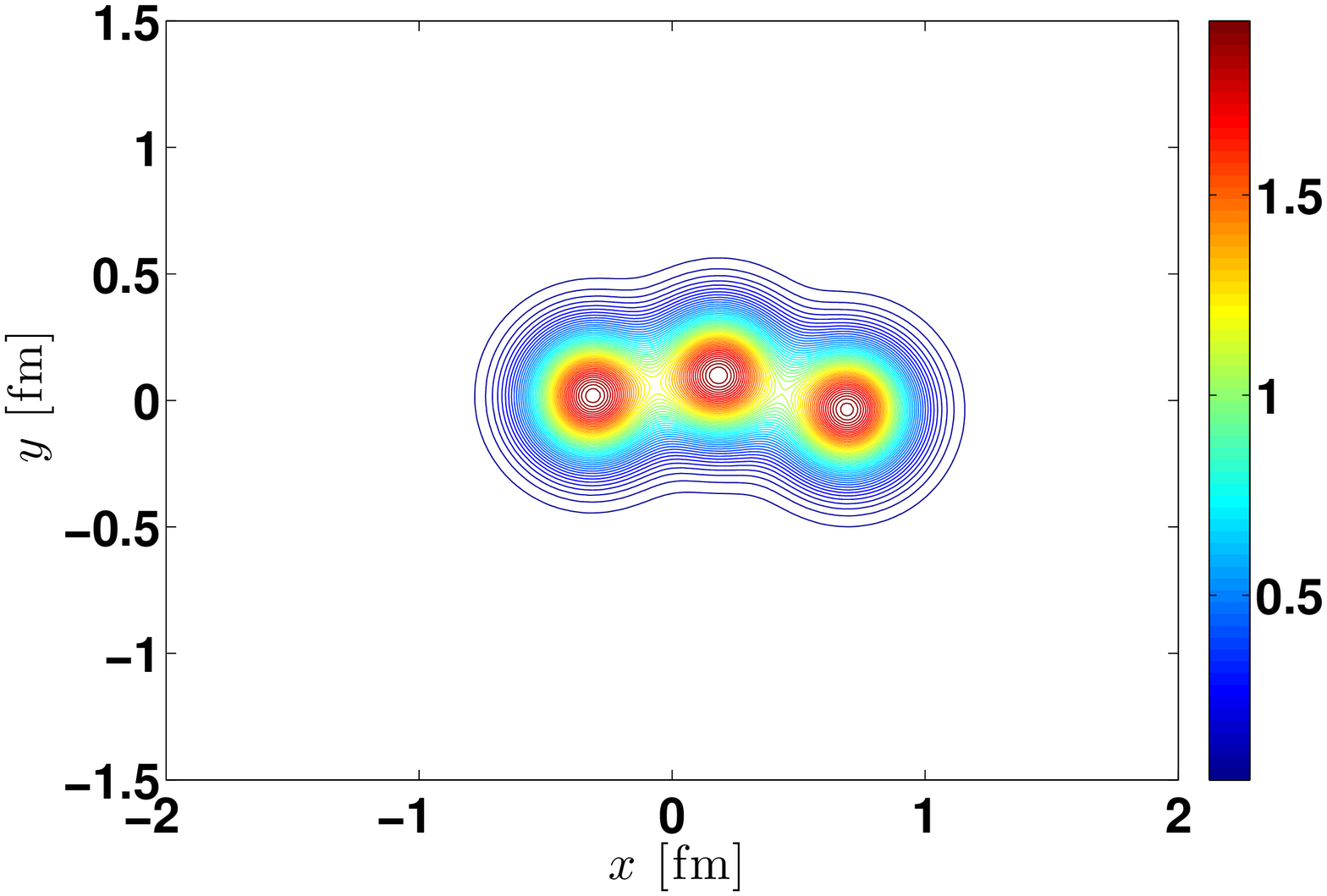}
\centering\includegraphics[width=\columnwidth,clip=true,angle=0]{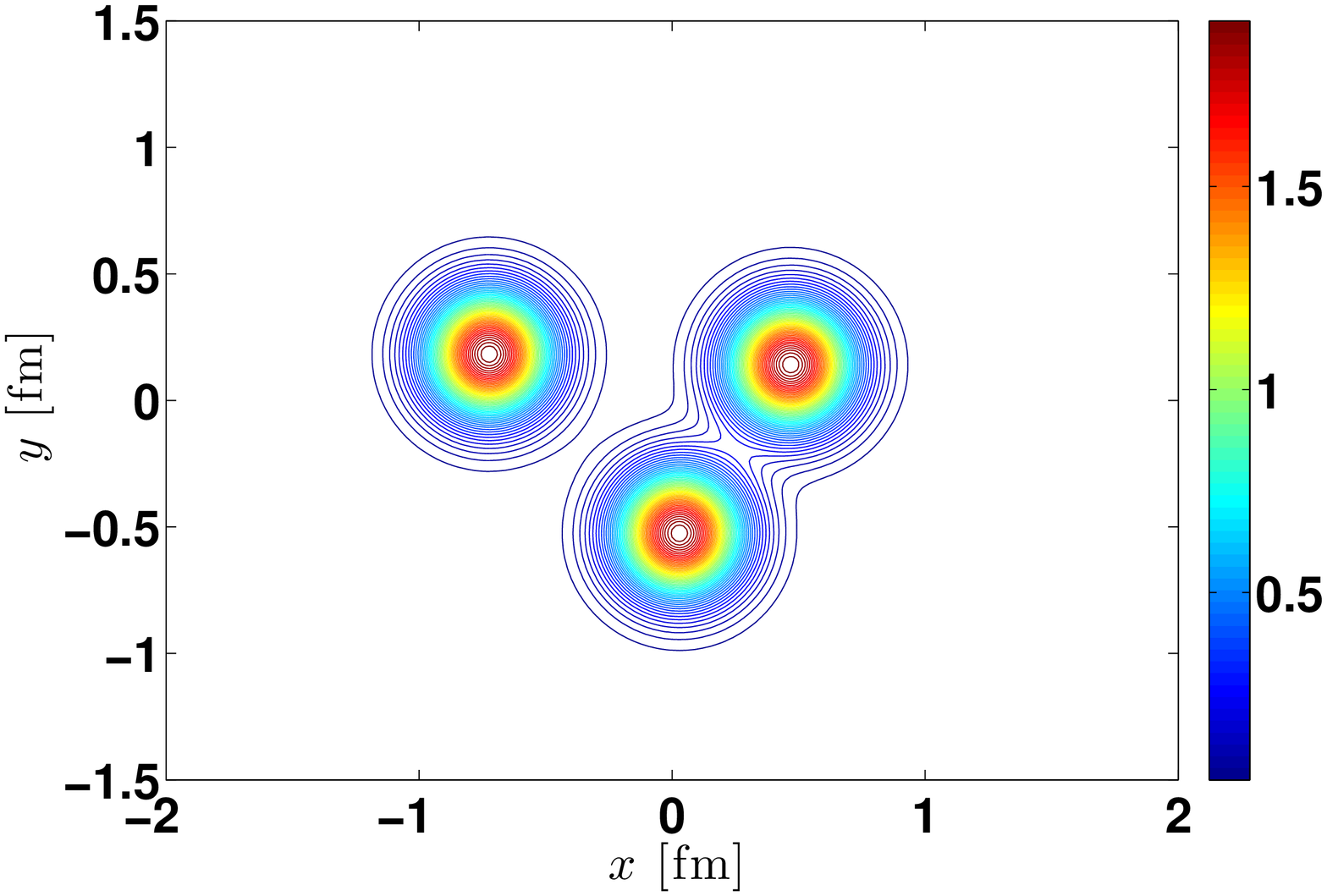}
\vspace{-1.0em}
\centering\includegraphics[width=\columnwidth,clip=true,angle=0]{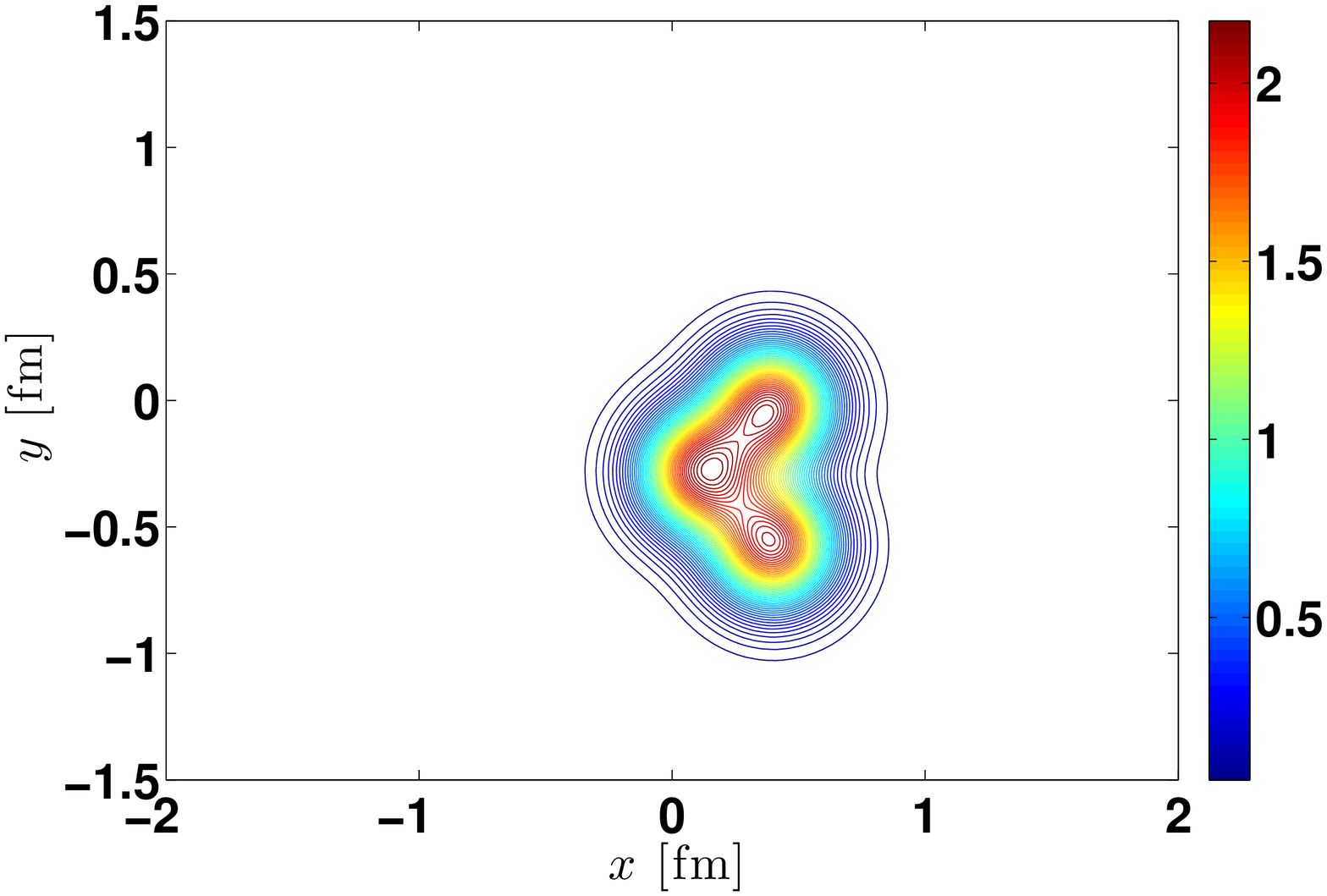}\caption{\small (color on line)  Examples of density profiles from Eqs.~(\ref{eq:T2ho}) with $B_q=0.7$ GeV$^{-2}$ characterizing the Gaussian shape of each gluon distribution around the constituent quark.}
\label{fig:Cshapes2ho}
\vspace{-1.0em}
\end{figure}

Gluon densities are obtained by adding, as was done earlier, around each constituent quark in the transverse plane as described by the profile (\ref{eq:T2ho}), a Gaussian gluon distribution with  parameter $B_q$. 
Examples of gluon transverse density profiles obtained in this way are shown in Fig.~\ref{fig:Cshapes2ho}. The results are analogous to those shown earlier in Fig.~\ref{fig:shapes1}, but in the present case the only free parameter is the width $B_q$ of the gluon cloud around each valence quark,  the positions of the quarks being  determined by the simplified 2 h.o. wave function and the electromagnetic sizes of proton and neutron (cf. Eqs.~(\ref{eq:rprn})), with no additional   free parameter. 

\section{\label{sec:Q_P} From Quarks to Partons}

The description of the quark states, within an appropriate quantum mechanical approach,  as detailed in the previous section, allows us to connect quarks and partons in a consistent way avoiding a new set of parameters entering the gluon distribution (cf. Eq.~(\ref{eq:xg})). In the present section we recall how to connect partons and quarks within a framework which makes use of  QCD perturbative evolution.

\subsection{Valence quarks and partons}
A simple description which connects the parton distributions to the momentum density of the constituents has been developed in the past (e.g. \cite{Traini_etal1997}). Within that approach the valence quark distribution for the bare nucleon is written as
\bea
q_V(x)\large|_{\rm bare} & = &{1 \over (1-x)^2}\,\int d^3k\, n(k)\,\delta \left({x \over 1-x} - {k_+ \over M}\right) = \nonumber \\
& = & 2 \pi {M \over (1-x)^2}\,\int_{k_m(x)}^\infty dk\,k\,n(k)\,,
\label{qV}
\eea
where
\be
k_m(x) = {M \over 2} \left|{x \over 1-x} - \left({m \over M} \right)^2 \,{1-x \over x}  \right|\,;
\ee
$k_+ = k_0 - k_z$ is the light-cone quark momentum fraction, $n(k)$ the quark momentum density distribution predicted by the specific QM wave functions, and $M$ and $m$ are the nucleon and constituent quark masses, respectively. One can check that $\int dx \,q_V (x, \mu_0^2) = \int d^3 k \,n(k) = {\cal N}_u + {\cal N}_d = 3$, i.e. the particle sum rule is preserved and the valence quark distributions (\ref{qV}) are defined within the correct support $0 < x < 1$.

In detail, the 2 h.o. quark momentum distribution within the proton reads
\bea
n(k) &=&  {\cal N}_u\,{1 \over \pi^{3/2}}\,{1 \over  \gamma_u^3}\, e^{-{k^2 \over \gamma_u^2}} + {\cal N}_d \,{1 \over \pi^{3/2}}\,{1 \over  \gamma_d^3}\, e^{-{k^2 \over \gamma_d^2}} = \nonumber \\ 
& \equiv & n_u(k)+n_d(k) 
\eea
where: ${\cal N}_u = 2$ and ${\cal N}_d=1$ are the numbers of $u$ and $d$ constituent quarks, while ${1 \over \gamma_u^2} = {3 \over 2}\,{4 \over 3\alpha^2+\beta^2}$ and ${1 \over \gamma_d^2} = {3 \over 2}\,{1 \over \beta^2}$ are the combinations of  parameters relevant for $u$ and $d$ momentum densities. One has:
\bea
u_V(x)\large|_{\rm bare} & = & {\cal N}_u \, {1 \over \sqrt \pi} {1 \over \gamma_u}\,{M \over (1-x)^2}\,e^{-{k^2_m(x) \over \gamma_u^2}}\,,\label{uV} \\
d_V(x)\large|_{\rm bare} &= & {\cal N}_d\, {1 \over \sqrt \pi} {1 \over \gamma_d}\,{M \over (1-x)^2}\,e^{-{k^2_m(x) \over \gamma_d^2}} \label{dV}\,; \\
q_V(x)\large|_{\rm bare} & = & \left[u_V(x)+d_V(x)\right]_{\rm bare}\,.
\label{qqV}
\eea
Of course the distributions (\ref{uV}) and (\ref{dV}) refer to an extremely low energy scale where the total amount of momentum is carried by the three valence quarks, with no gluon or sea contributions (bare nucleon).
In the next Section a concrete way to include the cloud degrees of freedom will be presented. 

\subsection{\label{sec:meson_gluon} From the meson cloud to sea quark and gluon distributions}

\begin{figure}[bp]
\centering\includegraphics[width=\columnwidth,clip=true,angle=0]
{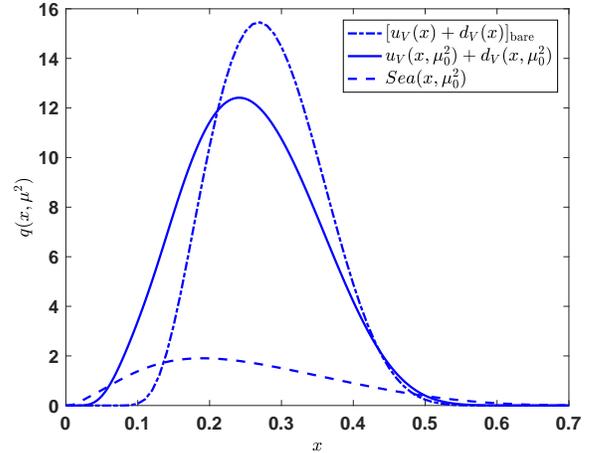}
\caption{\small (color on line) The parton distributions at the scale $\mu_0^2$ of the {\it physical} nucleon. The total non-perturbative sea due to Meson-Baryon fluctuations contains both strange and non-strange components (dashed line). The valence distribution (continuous line) is consistently renormalized (cf. Eqs.~(\ref{eq:partonicQ0}), (\ref{eq:Z})). For comparison also the valence distribution of the {\it bare} nucleon is shown (dot-dashed line).}
\label{fig:qxmu02}
\end{figure}

The quark model can be integrated with its virtual meson cloud incorporating $q \bar q$ pairs into the valence-quark picture of the parton distributions described in the previous sub-section, dressing the bare nucleon to a physical nucleon (see e.g. ref.~\cite{Traini2014} and references therein).
The physical nucleon state is built by expanding it [in the infinite momentum frame (IMF) and in the one-meson approximation] in a series involving bare nucleons and two-particle, meson-baryon (MB) virtual states.
The description of deep inelastic scattering (Sullivan process) implies that the virtual photon can hit either the bare proton $p$ or one of the constituents of the higher Fock states. In the IMF, where the constituent of the target can be assumed as free during the interaction, the contribution of those higher Fock states to the quark distribution of the physical proton can be written
\begin{eqnarray}
\delta q_p(x) & = &\sum_{BM} \left[ \int_x^1 \frac{dy} {y}\,f_{MB/p}(y)\,q_M\left(\frac {x}{y}\right) + \right. \nonumber \\
&& + \left. \int_x^1 \frac{dy} {y}\,f_{BM/p}(y)\,q_B\left(\frac {x}{y}\right)  \right]\,.
\label{eq:deltaqp}
\end{eqnarray}

The splitting functions $f_{BM/p}(y)$ and $f_{MB/p}(y)$ are the probability of the Fock state containing a virtual baryon (B) with longitudinal momentum $y$ and a meson (M) with longitudinal momentum fraction $1-y$.  The quark distributions in a physical proton are then given by
\be
q(x,\mu_0^2) = Z\,q_p^{\rm bare}(x) + \delta q_p(x)\,,
\label{eq:partonicQ0}
\ee
where $q_p^{\rm bare}$ is given by Eqs.~(\ref{uV}) (\ref{dV})  and $\delta q_p$ is from Eq.~(\ref{eq:deltaqp}).
\begin{equation}
Z = 1 - \sum_{MB}\,\int_0^1 dy\,f_{BM/p}(y)\,,\label{eq:Z}
\ee
is the renormalization constant and is equal to the probability to find the bare nucleon in the physical nucleon. In Fig.\ref{fig:qxmu02} the results are shown comparing the physical and bare parton distributions. The valence distribution is renormalized by the inclusion of the non-perturbative sea, the total sea distribution includes $\pi,\rho, \omega, K, K^*$ Meson-Baryon fluctuations, therefore the total sea is
$$
Sea(x,\mu_0^2) = 2\,\bar u(x,\mu_0^2) + 2\,\bar d(x,\mu_0^2) + s(x,\mu_0^2) + \bar s(x,\mu_0^2)\,,
$$
and strange and non-strange components are considered.

\begin{figure}[tbp]
\centering\includegraphics[width=\columnwidth,clip=true,angle=0]{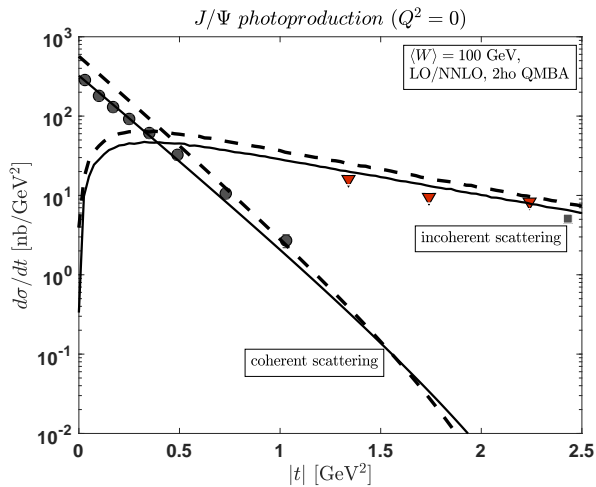}
\caption{\small (color on line)  Coherent and Incoherent photoproduction ($Q^2 = 0$) cross sections within the kinematical conditions of the HERA experiments ($\xP \approx 9.6 \cdot 10^{-4}$ for $\langle W\rangle = 100$ GeV).
The dashed lines  represent the QMBA-2ho predictions for the gluon distributions evolved at $LO$; the continuous lines represent results with gluons at $NNLO$ ($NLO$ and $NNLO$ predictions cannot be distinguished in the Figure, as emphasized in the text).
The incoherent scattering calculations within QMBA-2ho are made with $B_q = 0.7$ GeV$^{-2}$.
Data as in Fig.~\ref{fig:HERA_J_PSI_photo2}. }
\label{fig:LO-NNLO_coh_incoh_dXsec100_JPsi_Q2_0}
\end{figure}

The final results for the parton distributions at high resolution scale  $\mu^2 = \mu_0^2 + 4/r^2$ (cf. Eq.~(\ref{eq:mu})) are then obtained by evolving the initial distribution calculated at the scale $\mu_0^2$, by means of the DGLAP equations. More  details can be found in Ref.~\cite{Traini2014}. 
 \\

\section{\label{sec:2hoJ_Psi} $J/\Psi$ photoproduction within the quark model based approach (QMBA)}

Before showing the complete set of results for the $J/\Psi$ diffractive photoproduction in the coherent and incoherent channels, it is perhaps useful to  summarize the approach that we have presented in Sections~\ref{sec:QMBApproach} and~\ref{sec:Q_P}.
\begin{itemize}

\item[i)] We have proposed a generalization of the usual 
color-dipole picture (IPSat). The aim is to connect the diffractive scattering to proton properties like size, wave function symmetries, avoiding, as far as possible, {\it ad hoc} \\
parametrization like in Eqs.~(\ref{eq:TpGaussian}), (\ref{eq:mu}), and (\ref{eq:xg}).  We have constructed a  proton wave function in which the $SU(6)$ breaking is simply introduced by means of a two harmonic oscillator potential between constituent quarks \\ whose parameters are fixed by means of the experimental radii of neutron and proton. From that model the parton distributions are calculated at low resolution scale $\mu_0^2$ including a sea component by means of a well established formalism for the light-cone (perturbative) Meson - Baryon fluctuations. The procedure implies many parameters for the coupling constant, but they are taken from the most recent literature without any specific changes for the description of the diffractive scattering. Standard DGLAP evolution is applied to generate gluon distributions at the scale of the process, $\mu^2$. No further parameters are needed.

\item[ii)] The description of the {\it coherent} photoproduction of $J/\Psi$ does not need further ingredients and its calculation represent an absolute prediction  directly related to a  low-energy proton model. 
To describe {\it incoherent} diffraction an additional parameter ($B_q$) is needed to   relate the  fluctuations in the gluon density to the motion of the constituents quarks in the transverse plane, in analogy with Eqs.~(\ref{eq:Tbmodified}) and (\ref{eq:Tq}) as discussed in   Sect.~\ref{sec:2hoincoherent}. The parameter $B_q$  which controls the size of the gluon cloud around each valence quark,  is the only adjustable parameter of the approach. 

\end{itemize}

\subsection{\label{sec:2hocoherent} DGLAP evolution}

The predictions of the cross section for coherent and incoherent diffractive $J/\Psi$ photoproduction are compared with HERA data in 
Fig.~\ref{fig:LO-NNLO_coh_incoh_dXsec100_JPsi_Q2_0}. 
The leading order ($LO$) predictions (dashed lines) refer to a calculation where the gluon distribution that enters the dipole cross section is obtained by evolving the initial parton distribution using DGLAP equation at {\sl Leading Order}. The {\it next} or {\sl next-to-next to leading order} ($NNLO$) evolution equations are used for the calculation leading to the full lines (the difference between $NLO$ and $NNLO$ results could not be appreciated in the Figures). Strictly speaking, only the $LO$ calculation is fully consistent with the form of the dipole cross section that we use.
In the present calculation, the gluon distributions are predicted at high resolution scale starting from a low resolution physical picture of the nucleon. Their final values depend on the order of the evolution, which is reflected in the (weak) dependence of the diffractive cross sections  on the order of the evolution, as shown in Fig.~\ref{fig:LO-NNLO_coh_incoh_dXsec100_JPsi_Q2_0}. The fact that, as seen in Fig.~\ref{fig:LO-NNLO_coh_incoh_dXsec100_JPsi_Q2_0}, the experimental results appear to be better reproduced by the higher order evolution may reflect the better determination of the gluon distribution, although the slight inconsistency mentioned above prevents us to draw a too firm conclusion at this stage.  We may, minimally, regard the difference between the two sets of calculations as reflecting the intrinsic uncertainties of the theoretical model predictions here discussed.
In the following we will present results obtained at $NNLO$, mainly because they appear to be numerically more stable that the $LO$ ones.

\subsection{\label{sec:2hoincoherent} Fluctuations in incoherent scattering}

\begin{figure}[bp]
\centering\includegraphics[width=\columnwidth,clip=true,angle=0]
{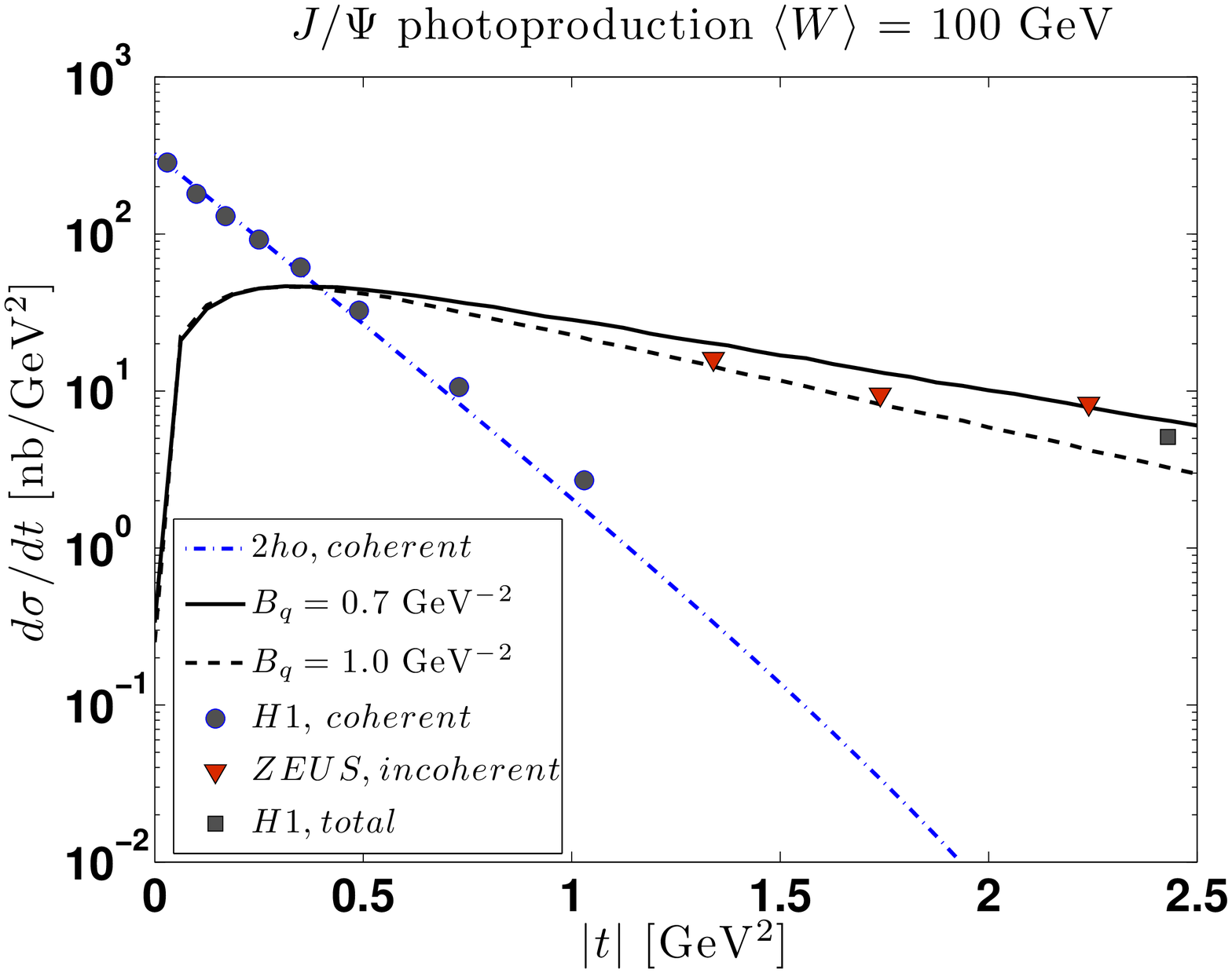}
\centering\includegraphics[width=\columnwidth,clip=true,angle=0]
{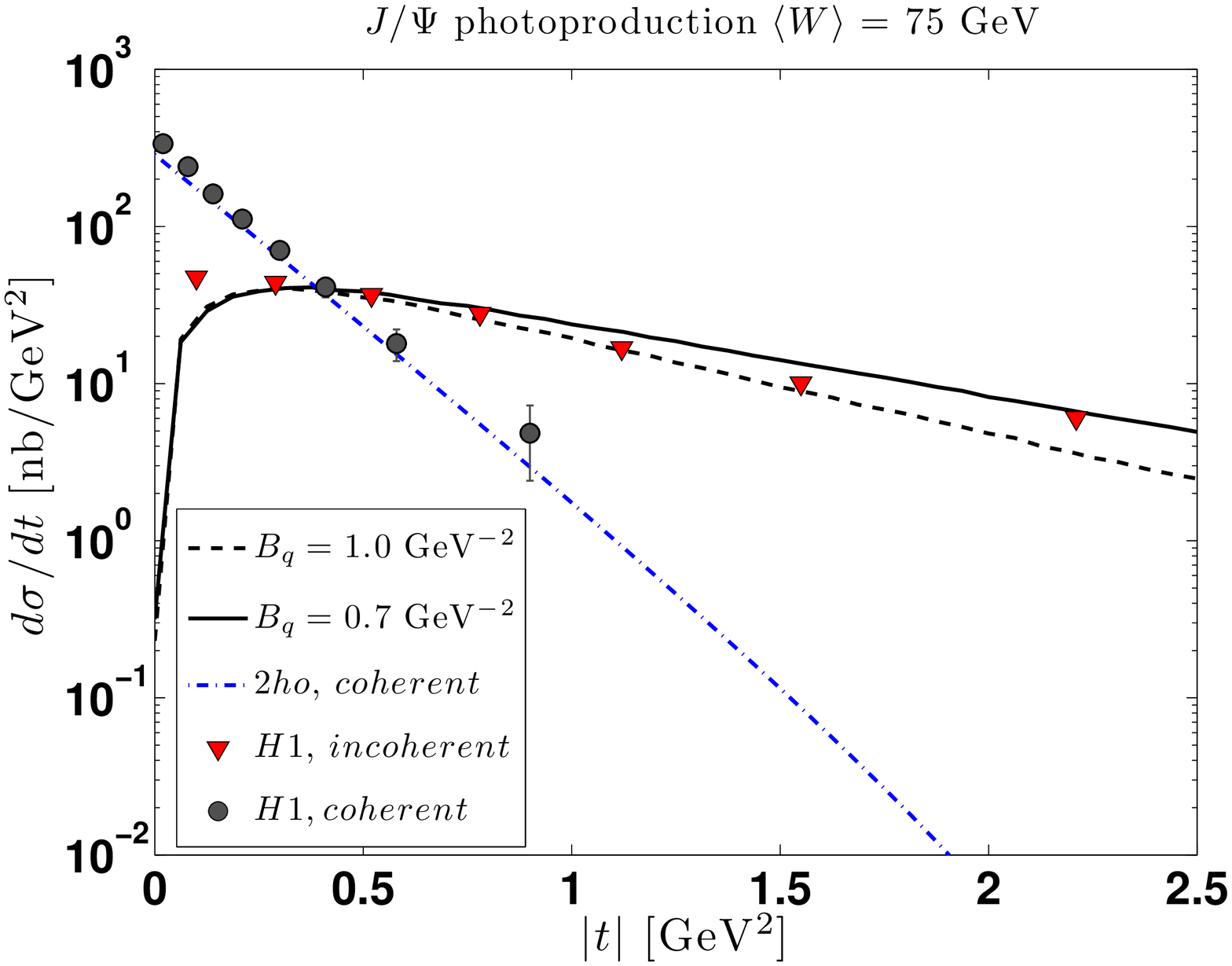}
\caption{\small Coherent and Incoherent photoproduction ($Q^2 = 0$) cross sections within the kinematical conditions of the HERA experiments: 
upper panel: $\xP \approx 9.6 \cdot 10^{-4}$ for $\langle W\rangle = 100$ GeV;
lower panel: $\xP \approx 1.7 \cdot 10^{-3}$ for $\langle W\rangle = 75$ GeV. 
The dot-dashed lines  represent the QMBA-2ho predictions for coherent scattering and the related $NNLO$ gluons.
Incoherent scattering calculations within QMBA-2ho are represented by the full lines and dashed lines according to the fluctuation parameter $B_q$ (see legend).
Data as in Fig.~\ref{fig:HERA_J_PSI_photo2}.}\label{fig:2hoFlW100_NNLO}
\end{figure}

Incoherent diffractive photoproduction can only be described by including gluon fluctuation effects, as we have emphasized earlier. The procedure to include gluon fluctuations extends that used with the Gaussian approximation for the profile functions (see Eqs.~(\ref{eq:Tbmodified}) and (\ref{eq:Tq}), and also Fig.~\ref{fig:HERA_J_PSI_photo2}). In the case of the QMBA profile,   the substitution analogous to Eq.~(\ref{eq:Tbmodified}) reads
\bea
&& T_{2ho}({\bf b}) \to {1 \over N_q} \sum_{i=1}^{N_q}T_{q}({\bf b}-{\bf b}_i), 
\label{eq:2hoTbTq1}\\
&& T_q({\bf b}) = {1 \over 2 \pi B_q}e^{-{\bf b}^2/(2 B_q)}.
\label{eq:2hoTbTq2}
\eea 
From a practical point of view one starts by considering  a sampling of the constituent quarks' positions (${\bf b}_i$, $i=1,2,3$, in the transverse plane), from the distribution 
$T_{2ho}({\bf b})$ of Eqs. (\ref{eq:2hoTbTq1}). This distribution includes part of the correlations between the quark positions coded in the   2 h.o. wave function. The gluon density around  each constituent quark  is  assumed to be Gaussian in the transverse plane, and is described by the function $T_q$ (\ref{eq:2hoTbTq2}).

For fixed $N_q$ ($N_q = 3$) the degree of fluctuations is controlled by the parameter $B_q$.
In Fig.~\ref{fig:Cshapes2ho} we have already shown an example of  lumpy proton configuration assuming $B_q = 0.7$ GeV$^{-2} = (0.1651\,{\rm fm})^2$ as suggested by the study of a Gaussian profile (Figs.~\ref{fig:shapes1} and related discussion).

Fig.~\ref{fig:2hoFlW100_NNLO} shows the relevant results for incoherent scattering comparing the calculations with the coherent component. The relevance of the fluctuations is confirmed, and also the value of the parameter $B_q$. The  Fig.~\ref{fig:2hoFlW100_NNLO} shows in fact that the values $0.7$ GeV $^{-2} \leq B_q \leq 1.0$ GeV$^{-2}$ remain the favorite range.
A consideration which is now independent from other parameters, specifically the parameter $B_{qc}$ needed to sample the position of the three quarks within the Gaussian approximation used in Ref.~\cite{M&Schenke2016}. In fact, in our quark model based approach, the quark positions are sampled directly from the  proton profile 
(\ref{eq:T2ho}) deduced from the quark model wave function. 
Of course the inclusion of a gluon distribution surrounding each valence quark does modify the global transverse gluon profile as already discussed in the case of a Gaussian transverse density (see Eq.~(\ref{eq:BqcBq}) and the related discussion). In the present case the r\^ole played by the parameter $B_{qc}$ of Sect.~\ref{sec:2hoincoherent} is assigned to the two parameters $B_u$ and $B_d$ of Eq.~(\ref{eq:T2ho}). In order to keep the transverse gluon root mean square radius fixed one has to replace (cfr. Eqs.(\ref{eq:BqcBq}), (\ref{eq:BuBd})).
\bea
B_u & \to & B_u - B_q; \nonumber \\
B_d & \to & B_d - B_q;
\eea
when $B_q > 0$. In that way the gluon rms radius
\be
\sqrt{\langle {\bf b}^2\rangle} = \sqrt{{2 \over 3} 2 (B_u +B_q) + {1 \over 3} 2 (B_d+B_q)} \approx 0.60\;{\rm fm}
\ee
will remain fixed varying $B_q$.

\subsection{\label{sec:roleofcorr}Quark correlations}

In order to illustrate the specific r\^ole of the $SU(6)$-breaking symmetry and the related quark correlations,  one can compare the results of the present QMBA 2h.o. correlated model with the limiting case of a single h.o. wave function which belongs to the $56$-th multiplet. The parameters of the two models are chosen in a consistent way, namely fixing the charge radius of the proton at the experimental value, cfr. Sect.~\ref{sec:T2ho}. Obviously the single harmonic oscillator model will predict a vanishing  charge radius of the neutron as discussed in the same Section, just because of the lack of $SU(6)$ configuration-mixing in the neutron wave function. The resulting transverse gluon 
profile function has been discussed in Sect.~\ref{sec:T2hob}. In particular, forcing the single harmonic oscillator model to reproduce the proton charge radius, will result in a rather large value of the transverse gluon root mean square radius. The corresponding coherent scattering cross section is, therefore, too low as it is evident from Fig.~\ref{fig:single_hoFlW100coh_incoh_NNLO}. The introduction of fluctuations does not alter the conclusion.

On the other hand, the incoherent scattering cross section calculated within the same single harmonic oscillator, $SU(6)$-symmetric, potential is able to follow the data behavior when a lumpy configuration is chosen ($B_q = 0.7$ GeV $^{-2}$, full line in 
Fig.~\ref{fig:single_hoFlW100coh_incoh_NNLO}). We could conclude that the incoherent scattering is so strongly dominated by the fluctuations that the r\^ole of quark correlations is unimportant. 

However such a conclusion needs to be qualified with the following considerations: 

\begin{itemize}

\item[i)] from Fig.~\ref{fig:2hoFlW100_NNLO}: if one uses a wavefunction which includes the proper correlation effects, fluctuations are essential to reproduce the incoherent cross section and, at the same time, the results are moderately sensitive to the free parameter $B_q$; 

\item[ii)] from Fig.~\ref{fig:single_hoFlW100coh_incoh_NNLO}: if one uses a wavefunction poorly correlated (e.g. a single harmonic oscillator), the effects of fluctuations are strongly sensitive to the parameter $B_q$.

\end{itemize}

The present calculations reveal therefore an interplay between the effects of correlations and those of fluctuations, the latter remaining however the crucial ingredient. 

\begin{figure}[tbp]
\centering\includegraphics[width=\columnwidth,clip=true,angle=0]
{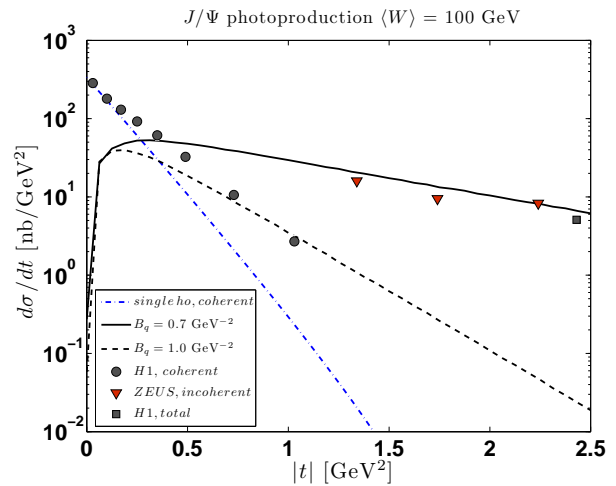}
\caption{\small (color on line) Coherent and Incoherent photoproduction ($Q^2 = 0$) cross section within the kinematical conditions of the HERA experiments ($\xP \approx 9.6 \cdot 10^{-4}$ for $\langle W\rangle = 100$ GeV).
The dot-dashed line reresents the single-ho predictions for coherent scattering.
Incoherent scattering calculations within single-ho are represented by the full and dashed lines according to the fluctuation parameter $B_q$ (see legend).
Data as in Fig.~\ref{fig:HERA_J_PSI_photo2}.}
\label{fig:single_hoFlW100coh_incoh_NNLO}
\end{figure}

\subsection{\label{sec:smallt} Small $|t|$ behavior}

The region at very small $|t|$ deserves a specific comment. Indeed this is the region where our predictions for incoherent scattering appear to deviate significantly from the data. When $|t|$ becomes small, the relevant fluctuations acquire a typical wavelength of the order of the size of the proton, and are not described by the geometrical fluctuations that we calculate. This can be seen from a simple analysis of Eq.~(\ref{eq:A_exclusive}) and Eq.~(\ref{eq:dXsqbarq}). In the limit where $\bmDelta\to 0$, the integral over the impact parameter in Eq.~(\ref{eq:A_exclusive}) becomes unconstrained, and the amplitude becomes proportional to the total dipole cross section, that is to the integral of Eq.~(\ref{eq:dXsqbarq}) over impact parameter. The reason why this kills the fluctuations can be easily understood by recalling how fluctuations are generated through the sampling of the valence quarks configurations, namely Eqs.~(\ref{eq:2hoTbTq1}), (\ref{eq:2hoTbTq2}): $T_{2ho}({\bf b})$ fluctuates because its value at a given ${\bf b}$ depends on whether there are valence quarks in the vicinity of ${\bf b}$, which is controlled by the function $T_q({\bf b}-{\bf b}_i)$. When $|{\bf b}|$ is constrained to be  small, i.e., $|{\bf b}|\leq R$ with $R$ the nucleon size, the final value of $T({\bf b})$ is sensitive to the location of the individual quarks and its value fluctuates. But when $|{\bf b}|$ is allowed to vary over distance larger than the nucleon size, which is the situation when $\bmDelta\to 0$,  the value of $T({\bf b})$ becomes insensitive to the precise location of the quarks. 

Thus the geometrical fluctuations of the kind discussed in the present paper are effective only at not too low momentum transfer.  In the region of small momentum transfer, extra sources of fluctuations are called for. This issue has been discussed in ref.~\cite{M&Schenke2016}. There, the authors have argued that fluctuations of the gluon density around each valence quark, that they express in terms of the fluctuations of the saturation momentum, can account for the missing ingredient, and can be tuned to reproduce the data in the small $|t|$ region.  Note that such fluctuations of the gluon density of the proton could also be understood in terms of the fluctuations of the size of the dipole going through the proton (see e.g. \cite{Blaizot:2016qgz} for a recent discussion of such issues). We have already indicated that such fluctuations are explicitly left out in the present calculation. Note also that the fluctuations of the dipole size appear to be the relevant ones  at $t=0$ in the approach based on cross section fluctuations, as discussed for instance recently in Ref.~\cite{Guzey:2018tlk}. The discussion of electroproduction in the next section will provide other indications on the importance of these fluctuations.

\begin{figure}[bp]
\centering\includegraphics[width=\columnwidth,clip=true,angle=0]
{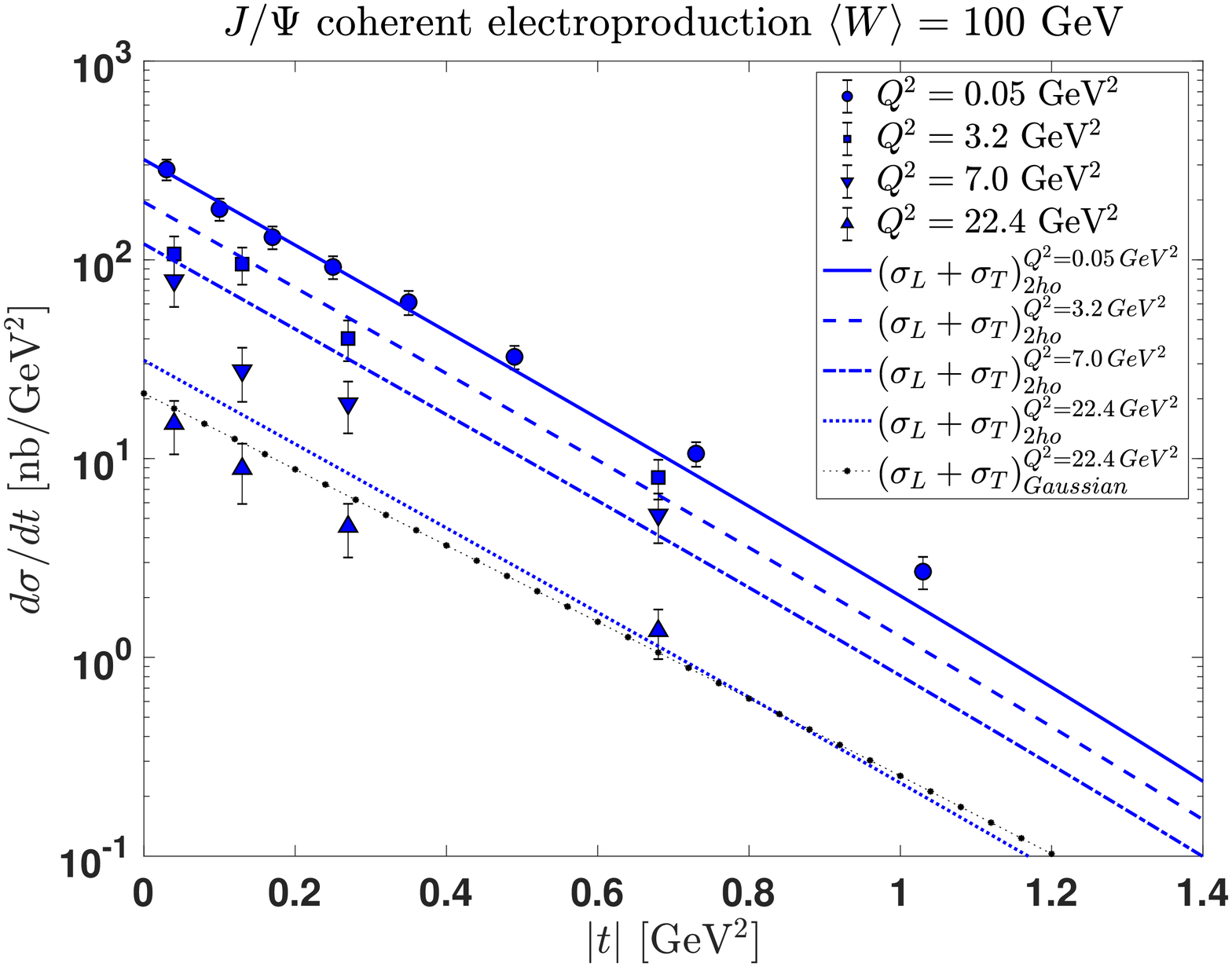}
\centering\includegraphics[width=\columnwidth,clip=true,angle=0]
{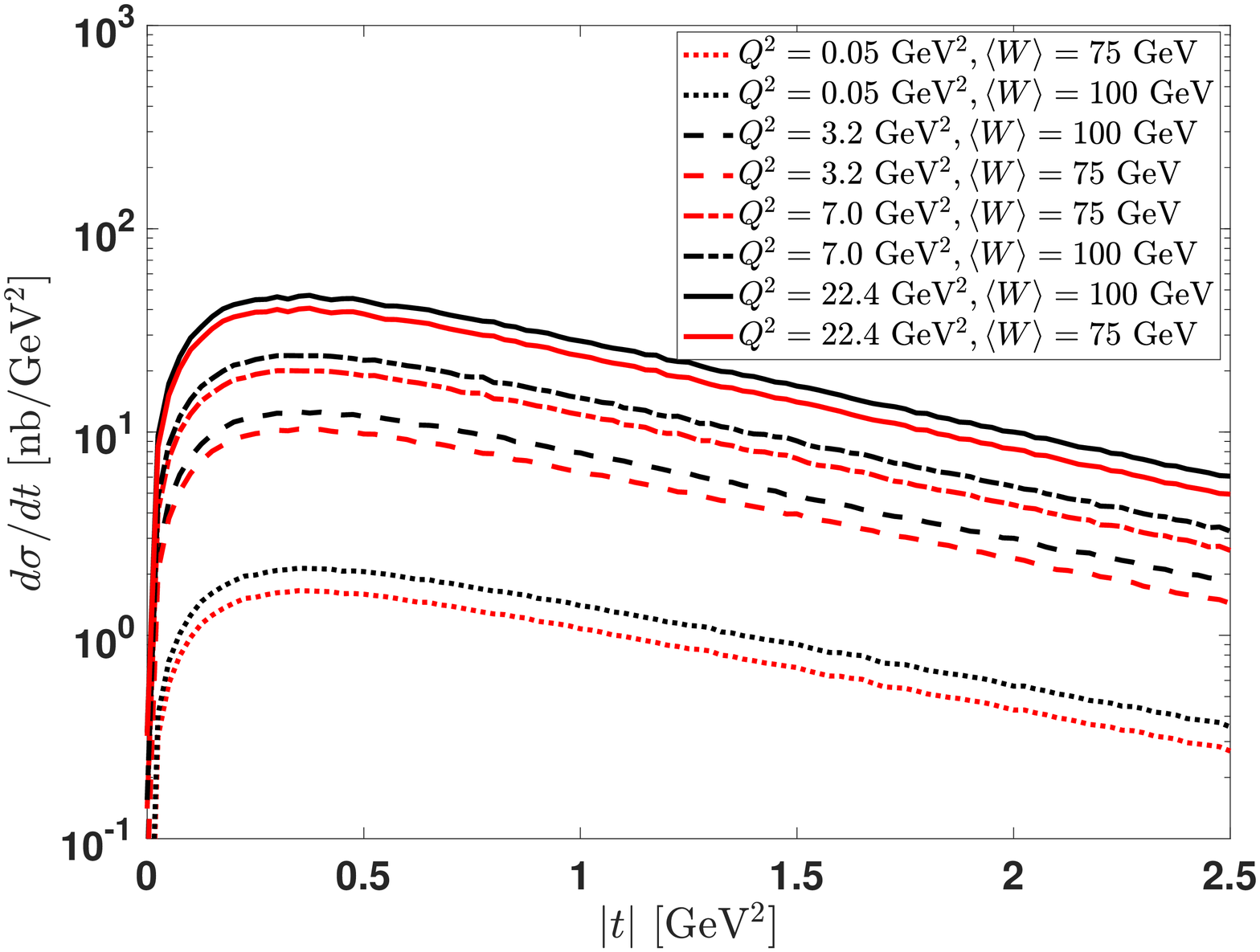}
\caption{\small (color on line)
{upper panel:} Coherent differential cross section ($\sigma_L + \sigma_T$) for electroproduction of $J/\Psi$ as a function of the total momentum transfer (square) $t$. Predictions within the quark-model-based approach (2 h.o.) and related $NNLO$ gluon profile are compared with data from HERA (H1) \cite{H1W100coh2006}. For $Q^2=22.4$ GeV$^2$ also the predictions of the Gaussian profile function ($B_G=4.0$ GeV$^{-2}$),  and the related ($LO$) gluon fit are shown as an example. 
{In the lower panel:} the incoherent components of the cross section at different $Q^2$ and within the QMBA (2 h.o.). The gluon fluctuations are included by using the same gluon distribution around each single quark used for $J/\Psi$ photoproduction ($B_q = 0.7$ GeV$^{-2}$).}
\label{fig:2hoW75_100_coh_incoh_Q2_NNLO}
\end{figure}

\section{\label{sec:electroproduction}$J/\Psi$, $\rho$ and $\phi$ electroproduction within the QMBA}

In the present section we complete the presentation of the QMBA approach to the kinematical conditions of electroproduction, i.e. for $Q^2 > 0$. Diffractive data exist for $J/\Psi$, and lighter vector mesons like  $\rho$ and $\phi$, and whenever possible, we compare our results to the available data. When appropriate, we also compare with the predictions based on the simple Gaussian profile function introduced in Sect ~\ref{sec:CoherentGaussian}.  The Boosted wave functions used for the vector mesons are described in~\ref{sec:forwardmeson}.

\subsection{\label{sec:eJPsi}$J/\Psi$ electroproduction}

A systematic comparison of our results with the HERA data is presented in the upper panel of Fig.~\ref{fig:2hoW75_100_coh_incoh_Q2_NNLO} for the $J/\Psi$ coherent electroproduction. The data are rather well reproduced within the QMBA description of the  transverse gluon shape (cf. Eq.~(\ref{eq:T2ho})) with no {\it ad hoc} parameters. The slopes of the curves are essentially determined by the geometrical size of the nucleon, fixed by the two parameters of the $2ho$ wave function (cf. Sec. \ref{sec:QMBApproach}), while the gluon distribution entering the dipole cross section~(\ref{eq:dXsqbarq})  keeps the form determined at $Q^2 = 0$, i.e. for $J/\Psi$ photoproduction (cf. Sect.~\ref{sec:meson_gluon}). The agreement deteriorates somewhat at large $Q^2$. For $Q^2 = 22.4$ GeV$^2$ the Gaussian-model  appears to perform slightly better.

The lower panel of Fig.~\ref{fig:2hoW75_100_coh_incoh_Q2_NNLO} provides predictions for the incoherent electroproduction, for which there are no available data. We have considered two kinematical conditions, $W=100$ GeV and $W=75$ GeV, and values of $Q^2$ that are identical to those of the data for the coherent scattering (upper panel). The geometrical fluctuations are calculated following the procedure discussed for the $J/\Psi$ photoproduction in the previous section. We recall that at low transfer, these predictions should not be trusted, for the reasons discussed  in Section \ref{sec:smallt}.

\begin{figure}[tbp]
\centering\includegraphics[width=\columnwidth,clip=true,angle=0]
{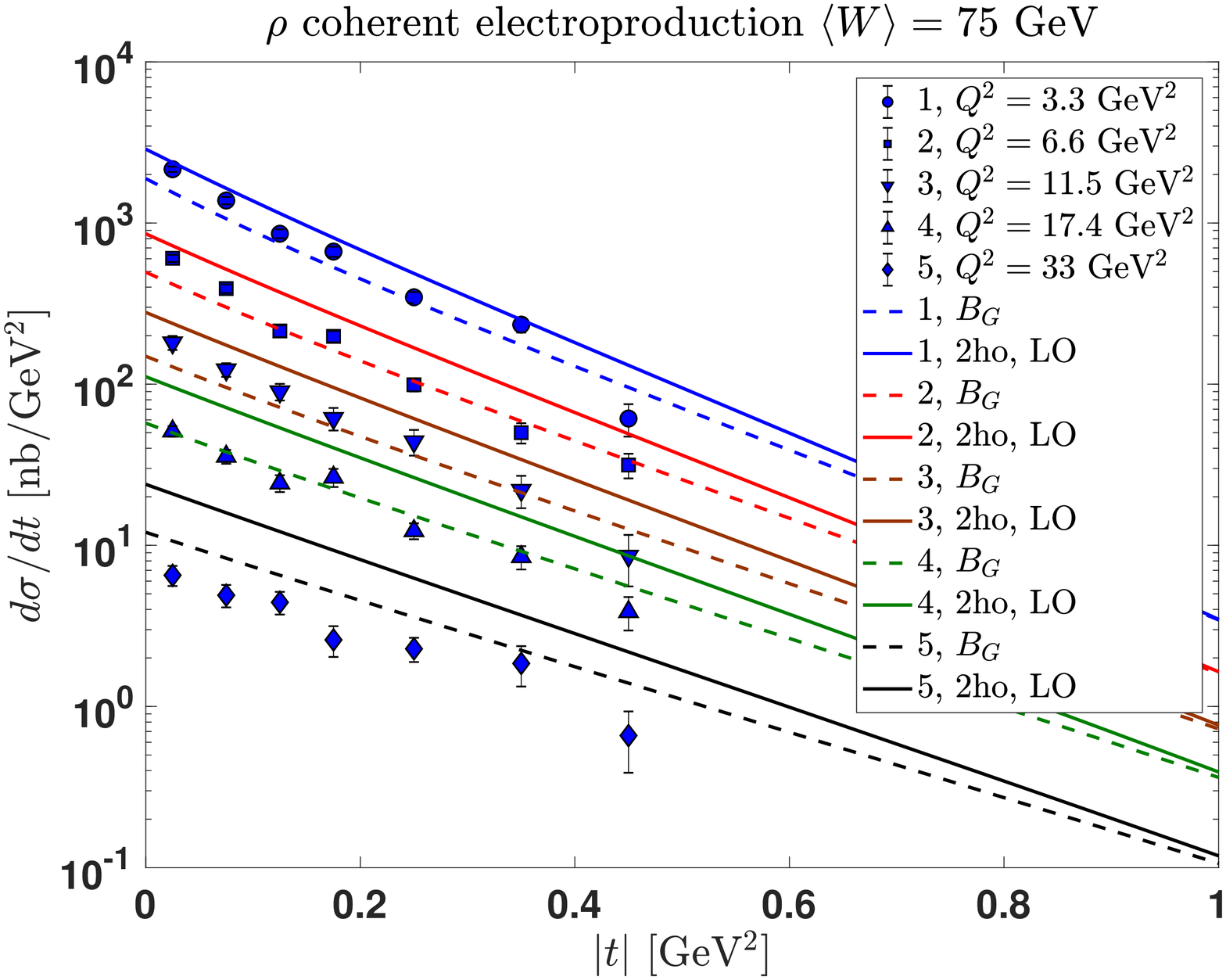}
\centering\includegraphics[width=\columnwidth,clip=true,angle=0]
{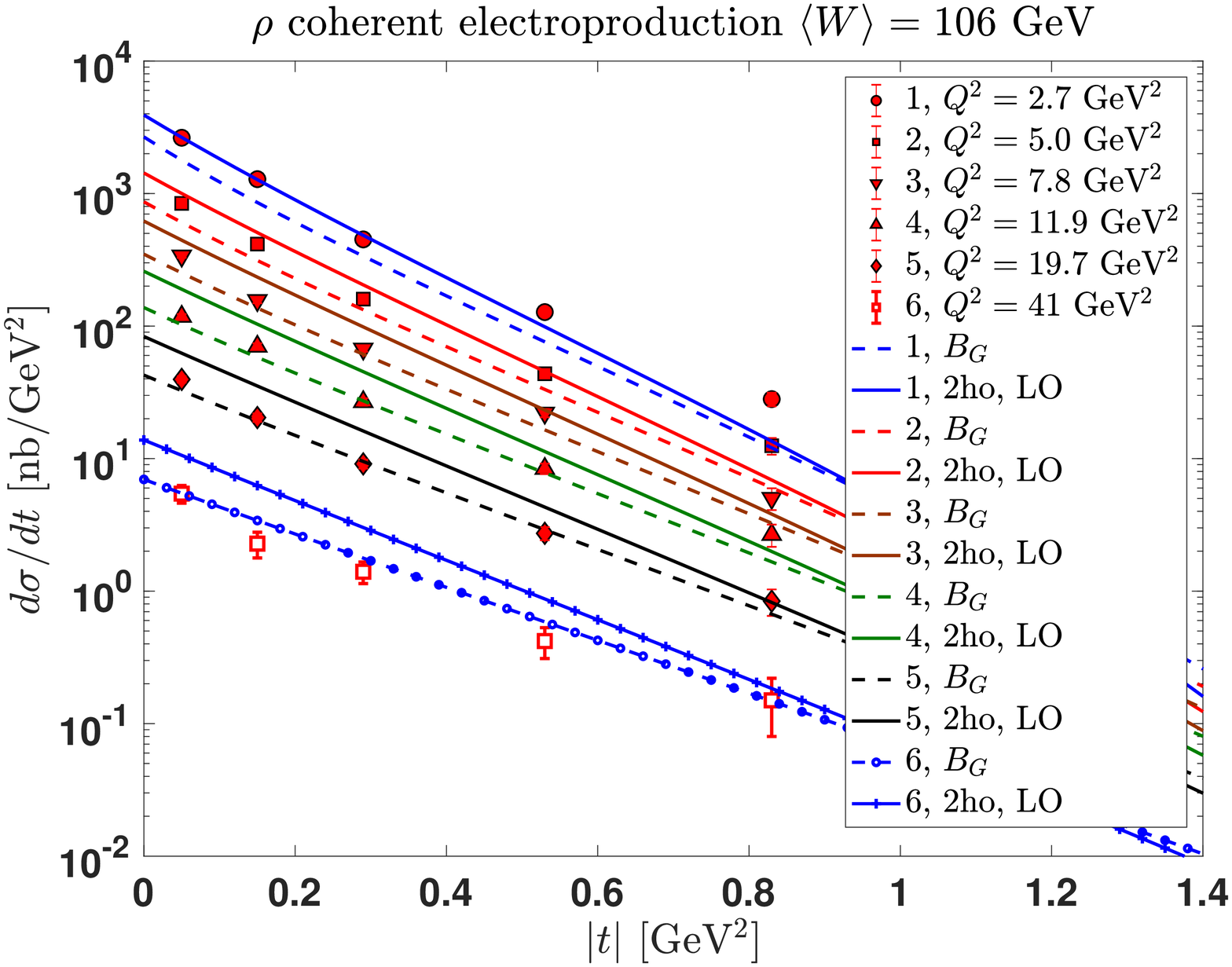}
\caption{\small (color on line) Differential cross sections for coherent electroproduction of $\rho$ mesons as a function of the total momentum transfer (square) $t$. The two set of data refer to the HERA experiments: H1 (upper panel) \cite{e-H1_rho_phi_2010}  and ZEUS (lower panel) \cite{e-ZEUS_rho0_2007}. The data are compared with present calculations within the QMBA (2 h.o.) and the Gaussian ($B_G$) approximation profiles.}
\label{fig:rho_coh_Q2}
\end{figure}

\subsection{\label{e-LighterMesons}Lighter meson electroproduction}

\subsubsection{\label{sec:rho}$\rho$ production}

A large amount of data exist both for coherent and incoherent diffractive electroproduction of  $\rho$-mesons. An example is shown in 
Fig.~\ref{fig:rho_coh_Q2} where the $H1$ and $ZEUS$ data, within a large range of $Q^2$ values, are compared with our predictions. The present 2ho QMBA approach and the  Gaussian approximation ($B_G = 4.0$ GeV$^{-2}$)  is confronted to  both data sets (lower and upper panels).  Slopes and $Q^2$ dependence are rather well reproduced except for the largest values of $Q^2$. As was observed already in the case of the $J/\Psi$, the Gaussian profile function seems to leads to a better agreement at large $Q^2$. We note however that as $Q^2$ increases, the photon wave functions probably become inaccurate (the $Q^2$ dependence of the whole cross section is entirely due to the $Q^2$-dependence of the photon wave function~(\ref{sec:forwardPhoton}); also,  we use a conservative value for the mass $m_f$ that enters the meson wave function, as it can be seen from table~(\ref{tab:UNO}) in~\ref{sec:overlapPSI}). Finally, we use here the $LO$ evolution in $\mu(r)$ (cf. Eq.(\ref{eq:mu})). All these factors could play a role and further studies would be needed to pin down precisely their respective effects. 

The analysis of incoherent scattering represents a novelty in the study of diffractive vector meson electroproduction\footnote{An initial analysis of incoherent diffractive electroproduction of  $\rho$ mesons has been proposed in 
ref.~\cite{M&Schenke2016}.} and in Fig.~\ref{fig:Bq_1p5_rho_incoh_Q2} we show our main results. Once again fluctuations are the crucial ingredient in order to have a non vanishing incoherent cross section.  In the present case, however, the slope can be reproduced with a rather larger fluctuation parameter, i.e. $B_q = 1.5$ GeV$^{-2}= (0.2417$  fm$)^2$. Keeping  instead the value  $B_q=0.7$ GeV$^{-2}= (0.1651\,{\rm fm})^2$ used for the $J/\Psi$ photo and electro-production, would lead to too much fluctuations. For the largest value of $Q^2$ the QMBA results overestimate the data values while the Gaussian approximation predictions ($B_{q_c} = (4-1.5)$~GeV$^{-2}$ and $B_q=1.5$ GeV$^{-2}$) are in better agreement. (We recall that $B_{q_c} + B_q = 4$ GeV$^{-2}$, in the Gaussian approximation cfr. Sect.\ref{sec:Gaussianfluctuations}).

\begin{figure}[tbp]
\centering\includegraphics[width=\columnwidth,clip=true,angle=0]
{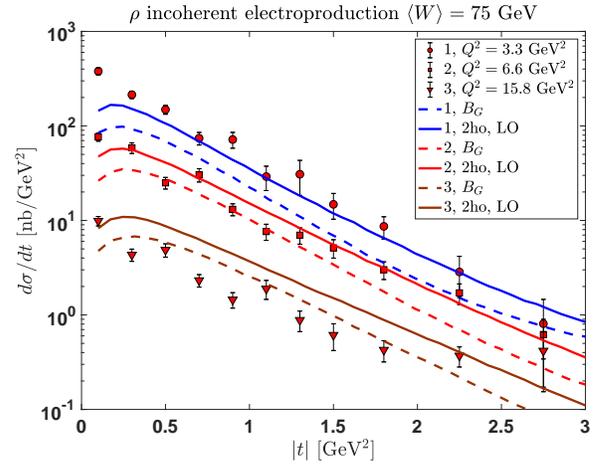}
%
\caption{\small (color on line) Differential cross sections for incoherent electroproduction of $\rho$ mesons as a function of the total momentum transfer (square) $t$. The data refer to the HERA H1 experiment \cite{e-H1_rho_phi_2010}  and are compared with present calculations within the QMBA (2 h.o.)  (full lines) and the Gaussian approximation ($B_G$) profiles (dashed lines). In both cases the fluctuation parameter $B_q = 1.5$ GeV$^{-2}$ and $LO$ gluon distributions are used.}
\label{fig:Bq_1p5_rho_incoh_Q2}
\end{figure}


\subsubsection{\label{sec:phi}$\phi$ production}

Results for coherent and incoherent $\phi$ elettroproduction are shown in Figs.~\ref{fig:phi_coh_incoh_Q2} and  \ref{fig:Bq_phi_incoh_Q2}. In particular in Fig.~\ref{fig:phi_coh_incoh_Q2} the coherent diffractive cross section is shown as evaluated within both the QMBA and the Gaussian profile. As in the case of the $\rho$ meson, both models fail in reproducing the  largest $Q^2$ data which seem to follow a different slope. 

\begin{figure}[bp]
\centering\includegraphics[width=\columnwidth,clip=true,angle=0]
{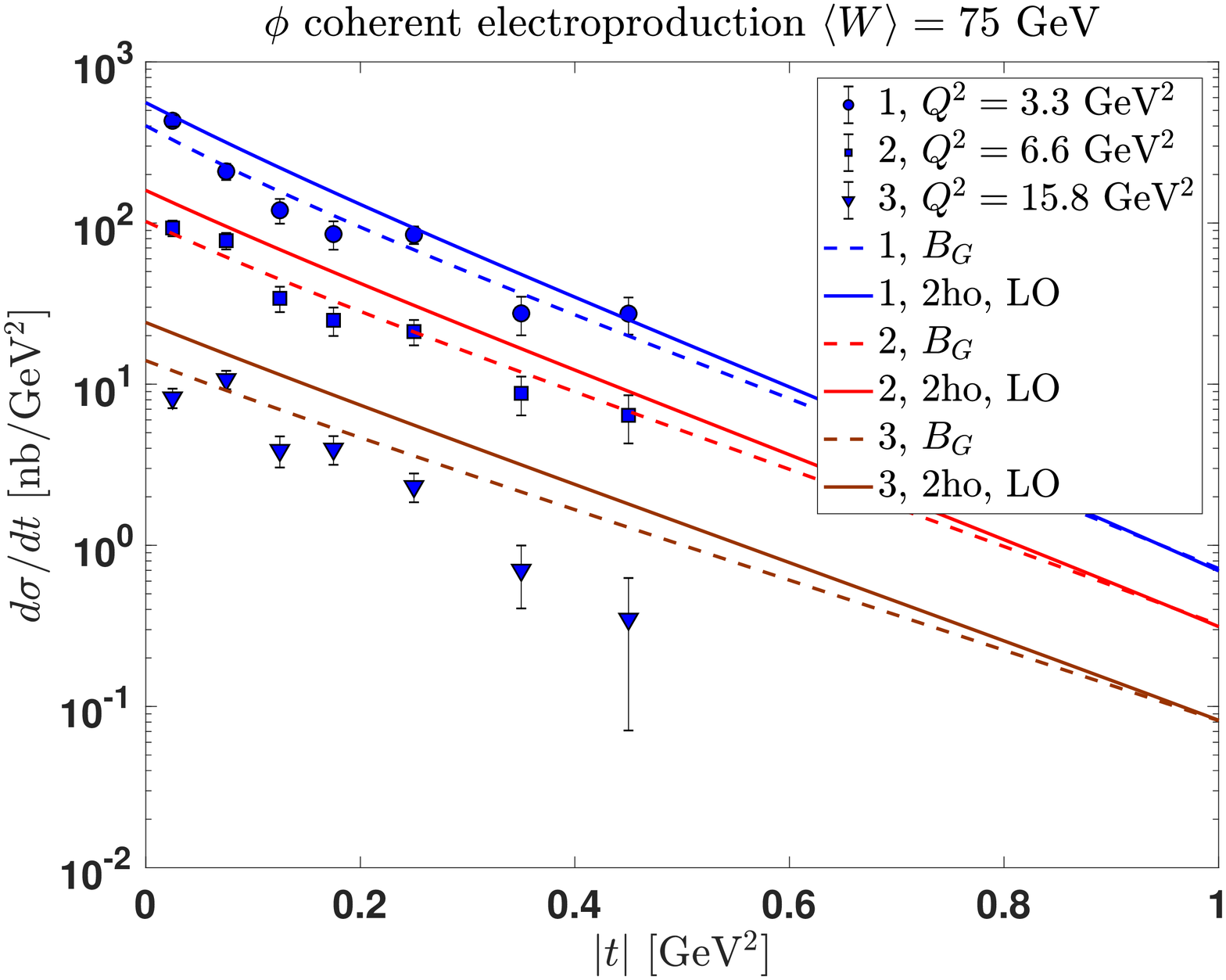}
%
\caption{\small (color on line) Differential cross sections for coherent electroproduction of $\phi$ mesons as a function of the total momentum transfer (square) $t$. The data refer to the HERA H1experiment  \cite{e-H1_rho_phi_2010}  and are compared with present calculations within the QMBA (2 h.o.) and the Gaussian ($B_G$) approximation profiles.
}
\label{fig:phi_coh_incoh_Q2}
\end{figure}

\begin{figure}[tbp]
\centering\includegraphics[width=\columnwidth,clip=true,angle=0]
{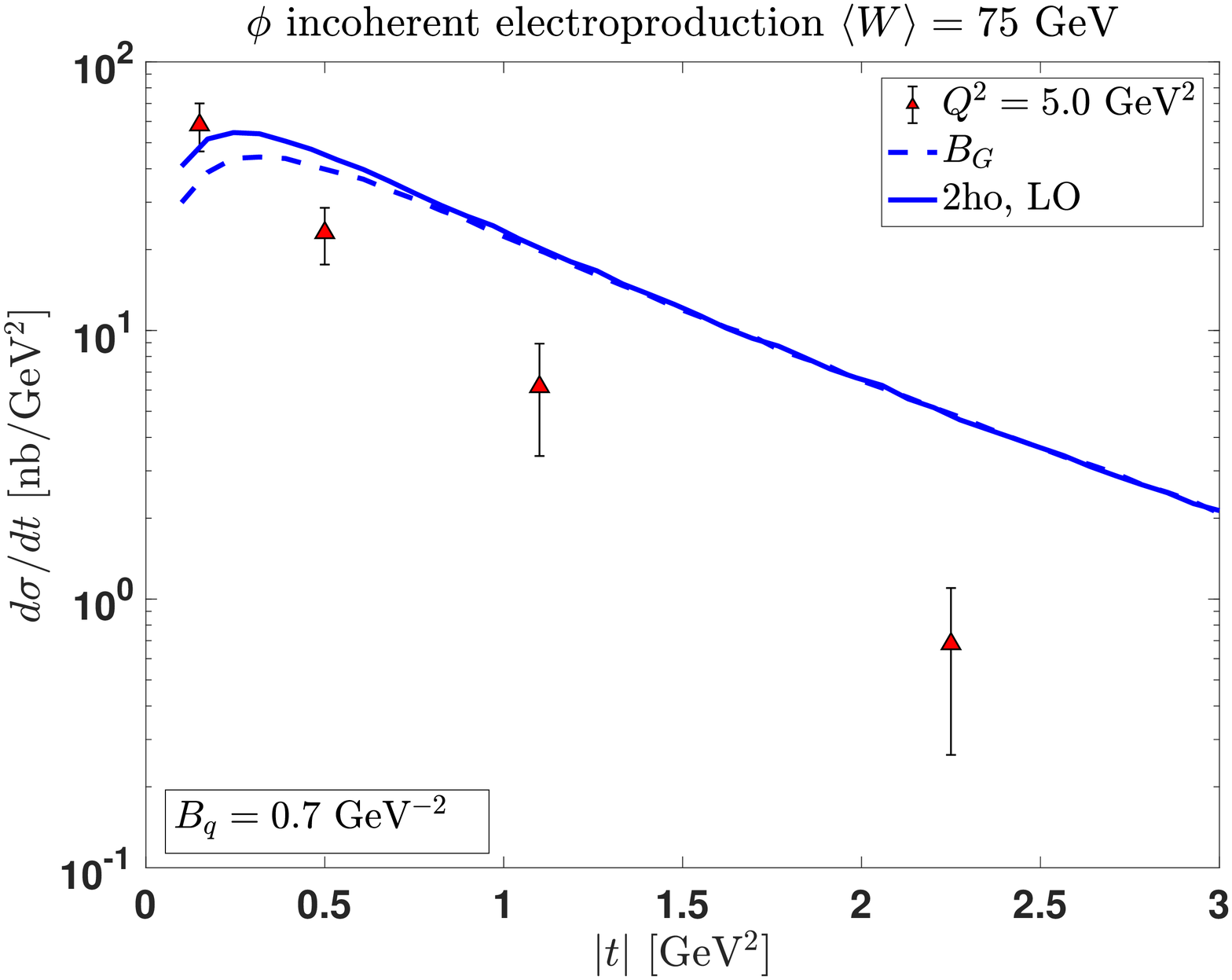}
\centering\includegraphics[width=\columnwidth,clip=true,angle=0]
{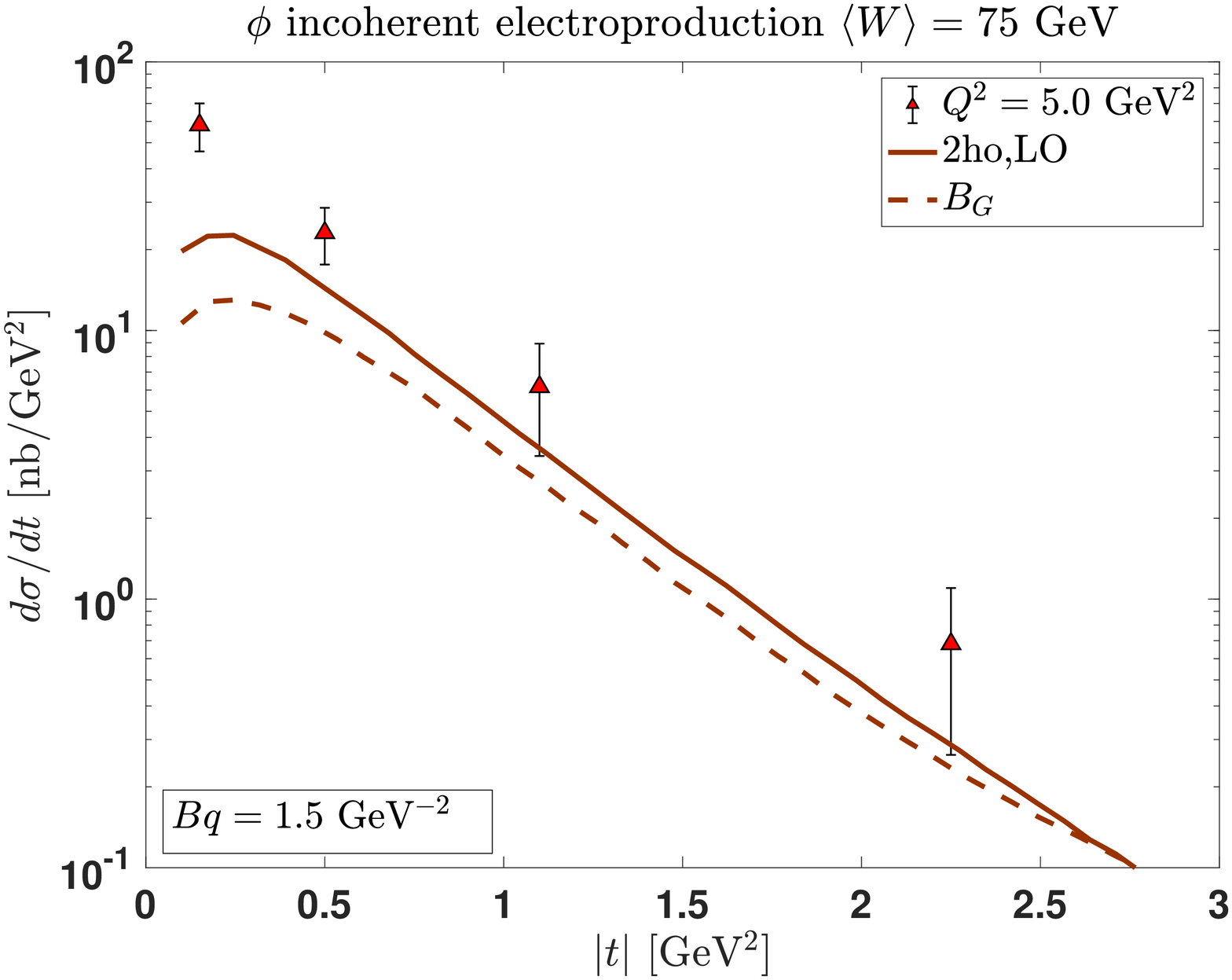}
%
\caption{\small  (color on line) Differential cross section for incoherent electroproduction of $\phi$ mesons as a function of the total momentum transfer (square) $t$. The data refer to the HERA H1 experiment \cite{e-H1_rho_phi_2010}  and are compared with present calculations within the QMBA (2 h.o.) (full lines) and the Gaussian ($B_G$) approximation profiles (dashed lines). In the upper panel the "lumpy" density profile selected to describe fluctuations in the $J/\Psi$ photo and electro production ($B_q = 0.7$ GeV$^{-2}$) is used and the predictions overestimate the data at large $|t|$; the lower panel results are obtained with a slightly "smoother" density profile ($B_q = 1.5$ GeV$^{-2}$) and are in good agreement with data. See text.}
\label{fig:Bq_phi_incoh_Q2}
\end{figure}

In order to emphasize the r\^ole of the fluctuation parameter $B_q$ we show
in the upper panel of Fig.~\ref{fig:Bq_phi_incoh_Q2}, the incoherent diffractive cross section 
evaluated within the QMBA and the Gaussian approximation fixing $B_q$ at the value used for the $J/\Psi$ production, i.e. $B_q=0.7$ GeV$^{-2}$. As in the case of $\rho$ meson production, for such value of the parameter $B_q$ one gets too much fluctuations. 
However, if one chooses the larger value already fixed in the case of diffractive production of the 
$\rho$, namely $B_q=1.5$ GeV$^{-2}$,  one obtains the results shown in the lower panel of Fig.~\ref{fig:Bq_phi_incoh_Q2} which are in better agreement with the experimental data.

Aside from the issues already  pointed out in our discussion of the $\rho$ meson electroproduction, it appears that  large dipoles are playing an important role for light mesons. Now, if this is the case, there are  features of our calculation that are not well treated. In particular, the fluctuations of the dipole size may induce additional fluctuations  that can in fact contribute to smear out the effect of geometrical fluctuations. Such a smearing is achieved here by increasing the size of the parameter $B_q$. Such fluctuations of the dipole size appear to be unimportant for heavy quarks, and at not too small values of $t$, and the $J/\Psi$ meson production is dominated by contributions of dipoles of small sizes. In this context, it would clearly be very interesting to have data on electroproduction of $J/\Psi$ mesons at various $Q^2$, to test for instance the predictions in Fig.~\ref{fig:2hoW75_100_coh_incoh_Q2_NNLO} and in view of new electron-ion collider (e.g. ref.~\cite{BondevaCepilaContreras2018}).

\section{\label{sec:conclusions}Conclusions and perspectives}

In the first part of the present work we have calculated the diffractive photoproduction of  $J/\Psi$, for both coherent and incoherent channels, including quark correlations in the evaluation of the gluon transverse density profile. The description of the gluon density in the transverse plane has been achieved through a  generalization of  the  IPSat model. This is based on an explicit quark model for the wave function of the valence quarks, with each constituent quark being surrounded by a gluon cloud.   Both spatial correlations, induced by a simple two-harmonic-oscillator potential model, and $SU(6)$-breaking symmetry correlations are included in the wave function, and these appear to play a r\^ole in the explicit calculation of the cross sections in the two channels. Since the parameters of the quark model are fixed on low energy properties of the proton and the neutron, no adjustable parameters are needed to calculate coherent diffractive production, and a single parameter needs to be selected to describe incoherent diffractive scattering: the width of the gluon distribution around each valence quark. The integrated gluon density is explicitly calculated from  the parton distribution deduced (at low resolution scale) from the quark model and evolved to the experimental, high energy,  scale using  DGLAP equations. A subtle interplay between quark correlations and geometric   fluctuations has been observed. \\

The second part of our work has been devoted to enlarge the domain of our study to diffractive vector meson production at $Q^2 > 0$, i.e. the electroproduction of $J/\Psi$ and lighter mesons like $\rho$ and $\phi$. Two new ingredients enter the calculations:  i) the $Q^2$ dependence of the cross sections; ii) the lighter mass of the mesons together with possible new non-perturbative effects.
\begin{itemize}

\item[i)] The $Q^2$ dependence of the cross sections is determined by the photon wave function, more precisely by the overlaps $(\Psi^*_V\Psi)_{T,L}(Q^2,{\bf r},z)$  of~\ref{sec:overlapPSI}.  The Gaussian Boosted wave functions show their limits in the descriptions of large dipoles as discussed in Sect.~\ref{e-LighterMesons}. The effect can be sizable at small as well as high $Q^2$ because of the interplay with the fluctuation of the dipole size. Calculations are in progress to model the lighter meson wave functions and dipole size fluctuations within a more elaborate approach better describing non-perturbative aspects (see also refs.\cite{AdS1,AdS2}).

\item[ii)] The smaller masses of lighter mesons introduce non-perturbative contamination. The net result is that the only parameter describing incoherent diffractive production in our approach (i.e. the size of the gluon cloud around each quark) differs from the heavy $J/\Psi$ meson from that  needed for lighter mesons like $\rho$ and $\phi$. This points to the relevance, for lighter systems, of fluctuations of different origin than the geometrical fluctuations discussed in this paper.  This is the case in particular of the  fluctuations in the dipole size, that we have argued could play also a role in the very small $t$ region. 

\end{itemize}

\noindent Finally, we note that the method introduced here can also be translated in the discussion of Deep Virtual Compton Scattering.

\begin{acknowledgements}
M.T. thanks the members of the Institut de Physique Th\'eorique, Universit\'e Paris-Saclay, for their warm hospitality during a visiting period when the present study was initiated. He thanks also the Physics Department of Valencia University for support and friendly hospitality. 
Useful remarks by  Heikki M\"antysaari are gratefully acknowledged. 
\end{acknowledgements}

\appendix

\section{\label{sec:overlapPSI} overlap functions $\left({\Psi^*\Psi_V}\right)_{T,L}(Q^2,{\bf r},z)$}

\subsection{\label{sec:forwardPhoton} forward photon wave function}

The forward photon wave function has been calculated perturbatively (e.g ref.~\cite{FSS2004})
\bea
&& \Psi_{h \bar h,\lambda = 0}(r,z,Q) =  e_f e \sqrt{N_c} \delta_{h,-\bar h}\,2\,Q\,z(1-z)\,{K_0(\epsilon r) \over 2 \pi}, \nonumber \\
&&{\rm for \, longitudinal} {\rm \,photon\,polarization\, (\lambda=0)};\\
\nonumber \\
&&\Psi_{h \bar h,\lambda = \pm}(r,z,Q)  =  \pm e_f e \sqrt{2 N_c}\left\{i e^{\pm i \theta_r}\left[ z \delta_{h,\pm}\delta_{\bar h,\mp} \right. \right. \nonumber \\
&& \left.\left. - (1-z) \delta_{h,\mp}\delta_{\bar h,\pm}\right]\partial_r + m_f \delta_{h,\pm}\delta_{\bar h,\mp}\right\} \nonumber \\
&& \times {K_0(\epsilon r) \over 2 \pi},\nonumber \\
&&{\rm for \, transverse} {\rm \, photon\,polarization\, (\lambda=\pm)},
\eea
where $r=|{\bf r}|$, $e=\sqrt{4 \pi \alpha_{\rm em}} \approx \sqrt{4 \pi /137}$, the subscripts $h$ and $\bar h$ are the helicities of the quark and the antiquark, respectively, $\theta_r$ is the azimuthal angle between the vector ${\bf r}$ and the $x$-axis in the transverse plane. $K_0$ is the modified Bessel function of second kind, $\epsilon^2 \equiv z(1-z)Q^2+m_f^2$ and $N_c = 3$ is the number of colors. The flavor dependence $f$ enters through the values of the quark charge $e_f$ and mass $m_f$, and $\partial_r K_0(\epsilon r) = - \epsilon K_1(\epsilon r)$.

\subsection{\label{sec:forwardmeson} forward vector meson wave function from ref.~\cite{IPSat2006}  (see also ref.~\cite{FSS2004})}

The simplest approach to modeling the vector meson wave function is to assume that the vector meson is predominantly a quark-antiquark state and that the spin and polarization structure is the same as in the photon case. The transversely polarized vector meson wave function (in complete analogy to the transverse polarized photon) is
\bea
\Psi^V_{h \bar h,\lambda=\pm}(r,z) & = & \pm \sqrt{2 N_c} {1 \over z(1-z)} \left\{i e^{\pm i \theta_r}\left[ z \delta_{h,\pm}\delta_{\bar h,\mp} \right. \right. \nonumber \\
&& \left.\left. - (1-z) \delta_{h,\mp}\delta_{\bar h,\pm}\right]\partial_r + m_f \delta_{h,\pm}\delta_{\bar h,\mp}\right\} \nonumber \\
&& \times \phi_T(r,z).
\eea
The longitudinally polarized wave function is slightly more complicated since the coupling  of the quarks to the meson, contrary to the photon case, is not local. One has:
\bea
\Psi^V_{h \bar h,\lambda=0}(r,z) & = & \sqrt{N_c} \delta_{h,-\bar h} \left[M_V + \delta \, {m^2_f -\nabla_r^2 \over M_V z(1-z)} \right] \phi_L(r,z),\nonumber \\
\eea
where $\nabla_r^2 \equiv (1/r) \partial_r + \partial^2_r$ and $M_V$ is the meson mass.
The nonlocal term was first introduced in refs.~\cite{NNPZ94_97_1,NNPZ94_97_2}.

\noindent The overlaps read then:

\bea
&& (\Psi^*_V\Psi)_T(Q^2,{\bf r},z) = {\hat e_f e \over \pi} {N_c \over z (1-z)}\left\{ m^2_f K_0(\epsilon r) \phi_T(r,z) \right.\nonumber \\
&& - \left. \left[z^2 + (1-z)^2 \right]\epsilon K_1(\epsilon r) \partial_r \phi_T\right\} 
\label{eq:overlapT}\\
&& (\Psi^*_V\Psi)_L (Q^2,{\bf r},z) = {\hat e_f e \over \pi} 2 N_c\,Q\,z (1-z) K_0(\epsilon r) \nonumber \\
&& \times \left[M_V \phi_L(r,z)  + \delta \, {m^2_f -\nabla_r^2 \over M_V z(1-z)} \phi_L(r,z)\right], 
\label{eq:overlapL}
\eea
where the effective charge $\hat e_f =2/3, 1/3$, or $1/\sqrt 2$, for $J/\Psi$, $\phi$, or $\rho$ mesons, respectively. In addition $\delta = 1$ is the natural choice done.

The boosted Gaussian wave functions in configuration space are written (see refs.~\cite{FSS2004,NNPZ94_97_1,NNPZ94_97_2})
\bea
\phi_{T,L}(r,z) & = &{\cal N}_{T,L}\,z(1-z) \exp \left[- {m^2_f {\cal R}^2 \over 8 z (1-z)} \right. \nonumber \\
&& - \left. {2 z (1-z) r^2 \over {\cal R}^2} + {m^2_f {\cal R}^2 \over 2} \right],
\label{eq:boostedG}
\eea
and ${\cal N}_{T,L}$ and ${\cal R}$ are fixed by normalization conditions and the decay width (see ref.~\cite{IPSat2006} for other details and Table~\ref{tab:UNO} for the values of the parameters).

\begin{table}[htp]
\caption{Parameters of the "boosted Gaussian" vector meson wave functions}
\begin{center}
\begin{tabular}{l c c c c c c c}
\hline
\hline
\\
Meson & & $M_V$/GeV & $m_f$/GeV & ${\cal N}_T$ & & ${\cal N}_L$ & ${\cal R}^2$/GeV$^{-2}$ \\
\\
$J/\Psi$ & & 3.097  & 1.4    & 0.578  & & 0.575  & 2.3  \\
\\
$\phi$   & & 1.019  & 0.14  & 0.919  &  & 0.825  & 11.2 \\
\\
$\rho$   &&  0.776  & 0.14  & 0.911  &  & 0.853  & 12.9 \\
\\
\hline
\hline
\end{tabular}
\end{center}
\label{tab:UNO}
\end{table}

\section{\label{sec:phen_corrections}Phenomenological corrections}

The derivation of the amplitude for the exclusive vector meson production (\ref{eq:A_exclusive})
(or DVCS amplitude if $V \to \gamma$, real photon) relies on the assumption that the S-matrix is purely real and, consequently, the exclusive amplitude ${\cal A}$ of  Eq.(\ref{eq:A_exclusive}) purely imaginary. The corrections due to the presence of the real part is accounted for  by the factor $(1+\beta^2)$ multiplying the differential cross sections (\ref{eq:dxsection_coh}), (\ref{eq:dxsection_incoh}). $\beta$ is the ratio of real and imaginary parts of the amplitude and is calculated by means of
\be
\beta = \tan {\pi \lambda \over 2}; \;\;\;\;\; {\rm with}\;\;\;\; \lambda \equiv {\partial \ln ({\rm Im }\,{\cal A}_{T,L}^{\gamma^* p \to V p}) \over \partial \ln(1/x_{\xP})}.
\ee
In addition for vector meson production (or DVCS) one should use the off-diagonal (or generalized) gluon distributions \cite{off-diagG}. Such a "skewed" effect is accounted for (in the limit of small $x_{\xP}$), by multiplying the gluon distribution $x g(x,\mu^2)$ by a factor $R_g$ given by \cite{IPSat2006}
\bea
R_g(\lambda_g) & = & {2^{\lambda_g+3} \over \sqrt \pi}{\Gamma(\lambda_g+5/2) \over \Gamma(\lambda +4)},
\nonumber \\
{\rm with}\;\;\;\;  \lambda_g & \equiv & {\partial \ln [x_{\xP} g(x_{\xP},\mu^2)]\over \partial \ln (1/x_{\xP})}.
\eea
The phenomenological corrections (in particular the  skewedness correction) are numerically relevant.
Evaluated without fluctuations and for $J/\Psi$ photoproduction, their (average) numerical values are around $10\%$ for the real part corrections and $40\%$ for the skewedness correction in the kinematical region $|t| \leq 0.5$ GeV$^2$ (see also ref.~\cite{M&Schenke2016}).


\begin{thebibliography}{9}

\bibitem{M&Schenke2016} {\it Revealing proton shape fluctuations with\\ 
incoherent diffraction at high energy},\\
Heikki M\"antysaari and Bj\"orn Schenke, \\
Phys. Rev. D 94, 034042 (2016).\\
arXiv:hep-ph/160701711

\bibitem{Miller:2007ri} {\it   Glauber modeling in high energy nuclear collisions},\\
 M.~L.~Miller, K.~Reygers, S.~J.~Sanders and P.~Steinberg, \\
Ann.\ Rev.\ Nucl.\ Part.\ Sci.\  {\bf 57}, 205 (2007),\\
ArXiv: nucl-ex/0701025

\bibitem{IPSat2002} {\it A modification of the saturation model: DGLAP evolution},\\
J. Bartels, K.J. Golec-Biernat, and H. Kowalski,\\
Phys Rev. D 66, 014001 (2002).\\
arXiv:hep-ph/0203258

\bibitem{IPSat2003} {\it Impact parameter dipole saturation model},\\
H. Kowalski and D. Teaney,\\
Phys. Rev. D 68, 114005 (2003).\\
arXiv:hep-ph/0304189

\bibitem{IPSat2006} {\it Exclusive diffractive processes at HERA within the dipole picture},\\
H. Kowalski, L. Motyka, G. Watt,\\
Phys. Rev. D 74, 074016 (2006).\\
arXiv:hep-ph/0606272

\bibitem{IPSat2013} {\it Analysis of combined HERA data in the impact-parameter dependent saturation model},\\
A.H. Rezaeian, M. Siddikov, M. Van de Klundert, and R. Venugopalan,\\
Phys. Rev. D 87, 034002 (2013).\\
arXiv:hep-ph/12122974

\bibitem{IPSat2017}{\it Energy dependence of dissociative $J/\Psi$ photoproduction as a signature of gluon saturation at LHC},\\
J. Cepila, J.G. Contreras and J.D. Tapia Takaki,\\
Phys. Lett. {\bf B766}, 186 (2017).\\
arXiv:hep-ph/160807559

\bibitem{DIS1}{\it Nuclear enhancement of universal dynamics of high parton densities},\\
H. Kowalski, T. Lappi, and R. Venugopalan, \\
Phys. Rev. Lett. {\bf 100}, 022303 (2008).\\
arXiv:hep-ph/07053047

\bibitem{DIS2}{\it Nuclear diffractive structure functions at high energies},\\
H. Kowalski, T. Lappi, C. Marquet, and R. Venugopalan, \\
Phys. Rev. C {\bf 78}, 045201 (2008).\\
arXiv:hep-ph/08054071

\bibitem{DIS2017}{\it Probing subnucleon scale fluctuations in ultraperipheral heavy ion collisions},\\
Heikki M\"antysaari and Bj\"orn Schenke,\\
Phys. Lett. {\bf B772} (2017) 832,\\
arXiv:hep-ph/170309256 

\bibitem{DIS3}{\it Coherent and incoherent $J/\Psi$ photonuclear production in a energy dependent hot-spot model},\\
J. Cepila, J. G. Contreras and M. Krelina,\\
Phys. Rev.  C {\bf 97}, 024901 (2018).\\
arXiv:hep-ph/171101855

\bibitem{M&Schenke2016PRL}{\it Evidence of strong proton shape fluctuations from incoherent diffraction},\\
Heikki M\"antysaari, Bj\"orn Schenke, \\
Phys. Rev. Lett. {\bf 117}, 052301 (2016).\\
arXiv:hep-ph/160304349

\bibitem{BP2002}{\it High-Energy Particle Diffraction}, \\
V. Barone and E. Predazzi,\\
vol.565 of {\it Texts and Monographs in Physics}. Springer-Verlag, Berlin Heidelberg, 2002.

\bibitem{gammaQED1} {\it Quantum Chromodynamics at High Energy}, \\
Y.V. Kovchegov and E. Levin,\\
Cambridge University Press, 2012.

\bibitem{Mueller1990}{\it  Small-$x$ behavior and parton saturation: \\
A QCD model},\\
A.H. Mueller,\\
Nucl. Phys. {\bf B335}, 115 (1990).

\bibitem{H1W100coh2006}{\it Elastic $J/\Psi$ production at HERA},\\
{\bf H1} collaboration, A. Aktas et. al., \\
Eur. Phys. J. {\bf C46}, 585 (2006).\\
 arXiv:hep-ex/0510016
 
\bibitem{H1W75coh2013}{\it Elastic and Proton-Dissociative Photoproduction of  $J/\Psi$ Mesons at HERA},\\
{\bf H1} collaboration, C. Alexa et. al., \\
Eur. Phys. J. {\bf C73}, 2466 (2013). \\ 
arXiv:hep-ex/1304.5162

\bibitem{ZEUSincoh2003} {\it Measurement of proton dissociative diffractive photoproduction of vector mesons at large momentum transfer at HERA},\\
{\bf ZEUS} collaboration, S. Chekanov et. al.,\\
Eur. Phys. J. {\bf C26}, 389 (2003).\\
arXiv:hep-ex/0205081

\bibitem{H1total2003}{\it Diffractive photoproduction of  $J/\Psi$ mesons with large momentum transfer at HERA}\,\\
{\bf H1} collaboration, A. Aktas et. al., \\ 
Phys. Lett. {\bf  B568}, 205 (2003).\\
arXiv:hep-ex/0306013

\bibitem{FSS2004}{\it Colour dipoles and $\rho$,$\phi$ electroproduction},\\
J.R. Forshaw, R. Sandapen, and F. Shaw,\\
Phys. Rev. D 69, 094013 (2004).\\
arXiv:hep-ph/0312172

\bibitem{NNPZ94_97_1}{\it Scanning the BFKL pomeron in elastic production of vector mesons at HERA},\\
J. Nemchik, N.N. Nikolaev, and B.G. Zakharov, \\
Phys. Lett. B 341, 228 (1994).\\
arXiv:hep-ph/9405355

\bibitem{NNPZ94_97_2}
{\it  Color dipole phenomenology of diffractive electroproduction of light vector mesons at HERA},\\
J. Nemchik, N.N. Nikolaev, E. Predazzi, and B.G. Zakharov, \\
Z. Phys. C 75, 71 (1997).\\
arXiv:hep-ph/9605231

\bibitem{OGEx} {\it Hadron masses in a gauge theory},\\
A. De R\'ujula, H. Georgi and S. Glashow, \\
Phys. Rev. D 12, 147  (1975).

\bibitem{GianniniSantopinto2015}{\it The hypercentral Constituent Model and its applications to baryon properties},\\
M.M. Giannini and E. Santopinto,\\
Chin. J. Phys. {\bf 53}, 020301 (2015).\\
arXiv:nucl-th/1501.03722

\bibitem{IsgurKarl1979} {\it $P$-wave baryons in the quark model},\\ 
Isgur and G. Karl, \\
Phys. Rev. D 18, 4187 (1978); \\
{\it Positive-parity excited baryons in a quark model with hyperfine interactions},\\ 
Phys. Rev. D 19, 2653 (1979).

\bibitem{ConciTraini1990} {\it  Quark Momentum Distribution in Nucleons},\\
L. Conci and M. Traini, \\
Few-Body Sistems, {\bf 8}, 123 (1990).

\bibitem{Franklin1968} {\it  A Model of Baryons Made of Quarks with Hidden Spin},\\
J. Franklin, \\
Phys. Rev. 172, 1807 (1968).

\bibitem{CapstickIsgur1986} {\it Baryons in a relativized quark model with chromodynamics},\\
S. Capstick and  N. Isgur, \\
Phys. Rev. D 34, 2809 (1986).

\bibitem{Kalosetal_2008}{\it Monte Carlo Methods}, \\
Malvin H. Kalos, Paula A. Whitlock, \\
2008, Wiley-VHC Verlag GmbH \& Co KGaA, Weinheim.

\bibitem{Traini_etal1997} 
{\it Deep Inelastic Parton Distributions and the Constituent Quark Model},\\ 
M. Traini, L. Conci and U. Moschella,\\
Nucl. Phys. {\bf A544},731 (1992). \\
{\it Constituent Quarks and Parton Distributions},\\
M. Traini,  V. Vento, A. Mair and A. Zambarda, \\
Nucl. Phys. {\bf A614}, 472 (1997).\\
{\it Quark Models and Meson Cloud in Deep Inelastic Scattering},\\
A. Mair and M. Traini,\\
Nucl. Phys. {\bf A628}, 296 (1998).\\
{\it Towards an Unified Picture of Constituent and Current Quarks},\\
S. Scopetta, V. Vento and M. Traini,\\
Phys. Lett. {\bf B421}, 64 (1998).\\
arXiv:hep-ph/9708262\\
{\it Polarized Structure Functions in a Constituent Quark Scenario},\\ 
S. Scopetta, V. Vento and M. Traini,\\
Phys. Lett. {\bf B442}, 28 (1998).\\ 
arXiv:hep-ph/9804302\\
{\it Polarized Parton Distributions and Light-Front Dynamics},\\ 
P. Faccioli, M. Traini and V. Vento,\\
Nucl. Phys. {\bf A656}, 400 (1999).\\
airXiv:hep-ph/9808201

\bibitem{Traini2014} {\it Next-to-next-to-leading-order nucleon parton distributions from a light-cone quark model dressed with its virtual meson cloud},\\ 
M. Traini, \\
Phys. Rev. D 89, 034021 (2014).\\
arXiv:hep-ph/13095814

\bibitem{Blaizot:2016qgz} {\it High gluon densities in heavy ion collisions},\\
J.~P.~Blaizot,\\  
Rept.\ Prog.\ Phys.\  {\bf 80}, no. 3, 032301 (2017).\\
arXiv:hep-ph/160704448

\bibitem{Guzey:2018tlk} {\it  Nucleon dissociation and incoherent $J/\psi$ photoproduction on nuclei in ion ultraperipheral collisions at the Large Hadron Collider},\\
V.~Guzey, M.~Strikman and M.~Zhalov,\\  
Phys. \ Rev.\ C 99, 015201 (2019).\\
arXiv:hep-ph/180800740

\bibitem{e-H1_rho_phi_2010}{\it Diffractive Electroproduction of $\rho$ and $\phi$ mesons at HERA},\\
{\bf H1} collaboration, F. Aaron \etal,\\
JHEP 1005 (2010) 032.\\
arXiv:hep-ex/09105831

\bibitem{e-ZEUS_rho0_2007}{\it Exclusive $\rho^0$ production in deep inelastic scattering at HERA },\\
{\bf H1} collaboration, S. Chekanov \etal,\\
PMC Phys. A1, 6  (2007).\\
arXiv:hep-ex/0781478

\bibitem{BondevaCepilaContreras2018}{\it Dissociative production of vector mesons at electron-ion colliders},\\
D. Bendova, J. Cepila, J. G. Contreras,\\
arXiv:hep-ph/181106479

\bibitem{off-diagG}{\it The effects of off diagonal parton distributions fixed by diagonal partons at small $x$ and $\xi$}\\
A.G. Shuvaev, K.J. Golec-Biernat, A.D. Martin, and M.G. Ryskin,\\
Phys. Rev. D 60, 014015 (1999).\\
arXiv:hep-ph/9902410

\bibitem{AdS1}{\it Diffractive $\rho$ and $\phi$ production at HERA using an AdS/QCD holographic light-front meson wavefunction},\\
Mohammad Ahmady, Ruben Sandapen, Neetika Sharma,\\
Phys. Rev. D 94, 074018 (2016).\\
arXiv:hep-ph/160507665

\bibitem{AdS2}{\it Vector Photoproduction using Holographic QCD},\\
Chang Hwan Lee, Hui-Young Ryu, Ismail Zahed,\\
Phys. Rev. D 98, 056006 (2018).\\
arXiv:hep-ph/180409300

\end{thebibliography}
\end{document}